\documentclass[superscriptaddress,aps,pra,twocolumn,showpacs,nofootinbib,longbibliography]{revtex4-1}
\usepackage{amsmath}
\usepackage{latexsym}
\usepackage{amssymb}
\usepackage[colorlinks=true,citecolor=blue,urlcolor=blue]{hyperref}
\usepackage{color}
\usepackage{graphics,epstopdf}
\usepackage{soul}
\usepackage[demo]{graphicx}
\usepackage{subfigure}

\begin{document}
\title{Protecting temporal correlations of two-qubit states using quantum channels with memory}
\author{Shounak Datta}
\email{shounak.datta@bose.res.in}
\affiliation{S. N. Bose National Centre for Basic Sciences, Block JD, Sector III, Salt Lake, Kolkata 700098, India}
\author{Shiladitya Mal}
\email{shiladitya.27@gmail.com}
\affiliation{S. N. Bose National Centre for Basic Sciences, Block JD, Sector III, Salt Lake, Kolkata 700098, India}
\affiliation{Harish-Chandra Research Institute, HBNI, Chhatnag Road, Jhunsi, Allahabad 211019, India}
\author{A. S. Majumdar}
\email{archan@bose.res.in}
\affiliation{S. N. Bose National Centre for Basic Sciences, Block JD, Sector III, Salt Lake, Kolkata 700098, India}

\date{\today}

\begin{abstract}
Quantum temporal correlations exhibited by violations of Leggett-Garg Inequality (LGI) and Temporal Steering Inequality (TSI) are in general found to be non-increasing under decoherence channels when probed on two-qubit pure entangled states. We  study the action of decoherence channels, such as amplitude damping, phase-damping and depolarising channels when  partial memory is introduced in a way such that two consecutive uses of the channels are time-correlated. We show that temporal correlations demonstrated by violations of the above temporal inequalities can be protected against decoherence using the effect of memory.
\end{abstract}


\maketitle

\section{Introduction}

Quantum correlations arguably are one of the most non-classical manifestations of quantum description of nature. Violations of local-realist description of physical laws by correlations between measurements performed locally on spatially separated entangled systems, are not only of foundational impact \citep{EPR, Bell, chsh}, but also enable many information processing tasks having advantage over their classical counterparts \citep{Bennett}.

The presence of quantum correlations beyond the microscopic domain, i.e., in case of large number of particles or large spins, has been studied \citep{mermin'90, roy'91, gm'83, peres'92} through  violations of local-realist inequalities. It has been shown that the amount of violation of the relevant local realist inequalities increase even in the limit of large numbers of particles and large spins considered together \citep{hm'95, cabello'02}.

Besides spatial correlations, sequential measurements performed on single systems yield temporal correlations. Non-classicality of such quantum correlations are revealed through violation of non-contextual \cite{ks'67} and macro-realist inequalities \citep{Anupam}.  Maximal temporal correlations in quantum mechanics may be obtained in a unified framework, as has been computed in \citep{budroni'13}.

Leggett and Garg\citep{Anupam} initially proposed a scheme for probing coherence of macroscopic systems employing violation of an inequality which is based on  two assumptions, namely, (i) {\it macrorealism per se}: at a given instant of time, a macroscopic object lies in a definite state among the accessibility of two or more macroscopically independent states, and (ii) {\it non-invasive measurability(NIM)}: it is possible to determine the state of a system without disturbing the state itself or its subsequent dynamics. Conjunction of these two assumptions together with induction (the future cannot determine the past) are known as macroscopic realism (MR). 

A testable algebraic consequence of the above assumptions is known as the Leggett-Garg inequality (LGI). Since then, various studies on LGI and their generalizations can be found in the literature\citep{b1, b2, b3, b4, b5, b6, b7, Mal, Mal1, b8}. Inequivalent necessary conditions of MR has been studied in \citep{nsit, udevi, saha}. Several experiments are also reported\citep{Expt, Expt1, Expt2, Expt3, Expt4} to test the Quantum Mechanical(QM) violations of LGI. Recently, LGI for two-qubit entangled states is proposed in \citep{lobejko} where the measurements are chosen to be Bell-state measurements(BSM) on the composite state of the system.

Later, the notion of temporal steering which is another way of probing non-classicality of temporal correlations, has been introduced in \citep{Chen, Li}. Analogous to the case of spatial steering, temporal steering has been linked to the issue of joint measureability\citep{Karthik}. An experimental demonstration of temporal steering has been reported\citep{Karol}. Further, a hierarchy between LGI violating correlations and temporal steering inequality (TSI) violating correlations has been proposed in \citep{hierar, hierar2}, analogous to the established hierarchy in spatial correlations.

Non-classical temporal correlations are potential candidates for various information processing tasks.  Contextuality is linked with quantum key distribution protocols \citep{cabello'11, arvind'17}. Temporal correlations violating macro-realism have been employed in information processing tasks such as quantum computation \citep{brukner'04}, device independent randomness generation \citep{mal'16}, secure key distribution \citep{shenoy'17}. Recently, it has been proposed that non-Markovianity of evolution may be detected by temporal steering  \citep{nonmarkov}. 
 
The applicability of quantum correlations in accomplishing various practical tasks is impacted by environmental interaction during the time evolution, which demolishes the possibility of probing non-classicality of correlations. Thus, the preservation of temporal quantum correlations against  environmental losses becomes very much important in order to perform  information-processing tasks. There are various protocols for preserving quantum spatial correlations, {\it viz}., quantum coherence\citep{Guo}, teleportation fidelity\citep{Pramanik}, entanglement\citep{Kim} and quantum steering\citep{Datta} against decoherence. Recently, Guo et al \citep{Yguo} have proposed a way to protect quantum correlations like  discord and  coherence using memory in noisy quantum channels.

The motivation of the present work is to propose a method for protecting temporal correlations against environmental loss. Some recent studies have computed the role of decoherence in temporal correlations\citep{Emary, hierar}.   In our present work, we first investigate the decay of temporal quantum correlations probed by the violations of LGI and TSI for two-qubit pure entangled states under uncorrelated quantum channels.  Then, by employing memory between two consecutive uses of quantum channels, we show that temporal correlations can be preserved against the diminishing effects of amplitude damping, phase damping, and depolarizing channels. We consider three scenarios of Bell-type measurements for calculating LGI  violation, and mutually unbiased measurements for calculating TSI violation.

The paper is organized in the following way. In Sec.\ref{2} we briefly discuss LGI and TSI. Next, in Sec.\ref{3} we characterise the correlated noise model including three noisy channels, {\it i.e.}, amplitude-damping, phase-damping and depolarizing channels. In Sec.\ref{4} and \ref{5} we discuss the effect of decoherence on LGI and TSI respectively, employing the effect of memory  in the consecutive use of channels. Finally, we conclude with some summarizing remarks in Sec.\ref{6}.

\section{Leggett-Garg inequality and Temporal steering inequality} \label{2}

\emph{Leggett-Garg inequality:} 
Consider sequential measurements on a single system evolving with time. In the first run, the  observable $Q$ is measured at time $ t_{1}$ and $t_{2}$, in the second at $t_{2}$ and $t_{3}$, and so on. The four-term LGI can be written as,
\begin{eqnarray}\label{lgi}
K_4 = C_{12} + C_{23} + C_{34} - C_{14} \leq 2,
\label{LGI}
\end{eqnarray}
where, the correlation function $C_{ij}=\langle Q(t_j) Q(t_i) \rangle =p^{++}(Q_{i},Q_{j})-p^{-+}(Q_{i},Q_{j})-p^{+-}(Q_{i},Q_{j})+p^{--}(Q_{i},Q_{j})$. Here, $p^{kl}(Q_{i},Q_{j})$ is the joint probability of getting outcomes `$k$' and `$l$' at times $t_{i}$ and $t_{j}$, respectively. In the macro-realist model which is also known as the non-invasive realist model the joint probability can be written in terms of some variable $\lambda$ which determines the value taken by the observable measured sequentially, given by
\begin{eqnarray}\label{nirm}
P(k,l| Q_{i}, Q_{j})=\sum_{\lambda}p(\lambda)p(k|Q_{i}, \lambda) p(l|Q_{j}, \lambda).
\end{eqnarray}
Violation of (\ref{lgi}) implies that there cannot be any non-invasive realist model for temporal correlations pertaining to the scenario.

\emph{Temporal steering inequality:} 
Temporal steering is the ability of preparing assemblage of states by measurements, which cannot be mimicked by some fixed ensemble of states (called hidden state ensemble, $\{p(\lambda), \rho_{\lambda}\}$, with $p(\lambda)$ being the distribution of states $\rho_{\lambda}$), without performing measurements. Suppose, in a single run the first measurement is done by Alice, and the second measurement is performed by Bob after receiving the system through some channel. In order to ascertain Alice's steerability, Bob needs to verify whether Alice gave him a post-measured assemblage. There exists a hidden state model for Bob when Alice is not capable of steering, and the joint probabilities can be written as
\begin{align}\label{ns}
P(Q_{i}=q_{i}, Q_{j}=q_{j})=\sum_{\lambda}p(\lambda)p(q_{i}|Q_{i}, \lambda) p^{Q}(q_{j}|Q_{j},\rho_{\lambda}).
\end{align}
Here, the superscript $Q$ on right hand side indicates that the probability emerges from quantum measurements according to the Born rule, $p(q_{i}|Q_{i}, \lambda)$ is the probability of outcome which Alice declares after performing her measurement, and $0\leq p(a_{k}|\lambda,k)\leq 1, \sum_{a}p(a_{k}|\lambda,k)=1$. Any inequality derived from this condition on joint probability is known as temporal steering inequality (TSI), the violation of which indicates steerability. One such inequality for $d$-dimensional systems has been introduced in \citep{Li}, given by
\begin{eqnarray}
S_{dU} = \sum_{i=1}^2 \sum_{a_i=0,b_{u(i)}=a_i}^{d-1} P(a_i,b_{u(i)}) < 1+\frac{1}{\sqrt{d}},
\label{TSI}
\end{eqnarray}
where, $P(a_i,b_{u(i)})$ is the joint probability of obtaining outcome $a_i$ on measuring $A_i$ at Alice's side at a given instant of time, and obtaining outcome $b_{u(i)}$ subject to Bob's measurement $B_i$ at some evolved time. Bob performs measurements in mutually unbiased basis pertaining to a $d$-dimensional system. QM maximum for the left hand side of Eq.(\ref{TSI}) is $2$ irrespective of dimensions.

\section{Correlated Noise Model} \label{3}

We briefly discuss the action of quantum noisy channels with memory.  In the case of a memoryless channel, the environmental correlation time is smaller than the time between two consecutive uses of the channel over two qubits. Therefore, the environmental back action can be ignored. Conversely, in the case of a memory channel, the environmental correlation time is greater than that between two consecutive uses of the channel. Suppose, between two consecutive uses of a channel $\xi$, each channel input acts independently. Then the entire channel action can be denoted as $\xi_2 = \xi \otimes \xi$, which implies memoryless quantum channels. On the other hand, a channel is time-correlated or a memory channel if it exhibits $\xi_2 \neq \xi \otimes \xi$.

A quantum channel can be defined as a completely-positive, trace-preserving map between two density matrices, and according to superoperator formalism, one can write the action of a channel on a quantum state, $\varrho$ as,
\begin{eqnarray}
\xi(\varrho)=\sum_i \mathcal{E}_i \varrho \mathcal{E}_i^{\dagger}
\end{eqnarray}
Decoherence channels map to output states which have less coherence than input states, and can be represented in terms of non-unitary matrices $\mathcal{E}_i$ called Kraus operators of the channel. $\mathcal{E}_i$'s are obtained by tracing out the environment from the global unitary operation acted upon both the system and the environment. Subsequently, the Kraus operators for two-qubit input state, $\rho$ can be represented as,
\begin{eqnarray}
\mathcal{E}_{ij} = \mathcal{E}_i \otimes \mathcal{E}_j = \sqrt{\mathcal{P}_{ij}} \mathcal{A}_i \otimes \mathcal{A}_j
\end{eqnarray}
where $\mathcal{P}_{ij}$ has the interpretation of being the probability of random sequence of operations applied to two qubits which are transmitted through the channel, and it satisfies the completeness relation $\sum_{i,j} \mathcal{P}_{ij} = 1$. If the channel is memoryless, then $\mathcal{P}_{ij} = \mathcal{P}_{i} \mathcal{P}_{j}$ and single qubit Kraus operators are independent of each other, whereas, for a memory channel, we have, $\mathcal{P}_{ij} = \mathcal{P}_{i} \mathcal{P}_{j|i}$ according to Bayes rule, where $\mathcal{P}_{j|i}$ is the conditional probability of the next operation following a Markov chain after performing the previous operation.

Two consecutive uses of a two-qubit quantum channel with partial memory can be encompassed using the Kraus operator,
\begin{eqnarray}
\mathcal{E}_{ij} = \sqrt{\mathcal{P}_i [(1-\mu) \mathcal{P}_j + \mu \delta_{ij}]} \mathcal{A}_i \otimes \mathcal{A}_j
\end{eqnarray}
where, the conditional probability function, i.e. $\mathcal{P}_{j|i}=(1-\mu) \mathcal{P}_j + \mu \delta_{ij}$, depends on the memory co-efficient of the channel,$\mu$ ($0\leq \mu \leq 1$). The Kraus operator formalism enables us to write the output $\xi_2(\rho)$ of an initial state $\rho$ transmitted through a two-qubit channel correlated with time-dependent Markov noise which is given by\citep{Macchiavello},
\begin{eqnarray}
\xi_2(\rho) = (1-\mu) \sum_{i,j=0,1} \mathcal{E}_{ij}~ \rho ~\mathcal{E}_{ij}^{\dagger} + \mu \sum_{k=0,1} \mathcal{E}_{kk}~ \rho ~\mathcal{E}_{kk}^{\dagger}
\label{mem}
\end{eqnarray}
where $\mu$=0 implies zero memory, and $\mu$=1 implies perfect memory. The two channels are correlated with probability $\mu$.

\subsection{Amplitude damping channel with memory}

Energy dissipation of a two-level quantum system by spontaneous emission of a photon into the vacuum is characterised by the amplitude damping channel. The Kraus operators for a single qubit are given by,
\begin{eqnarray}
\mathcal{E}_0 = \begin{pmatrix}
\sqrt{1-p} & 0\\
0 & 1
\end{pmatrix}; ~~~~~~~~
\mathcal{E}_1 = \begin{pmatrix}
0 & 0\\
\sqrt{p} & 0
\end{pmatrix}
\end{eqnarray}
where, the damping parameter $p$ ranges from $0$ to $1$.
Evolution of two-qubits under amplitude damping  without memory can be represented  by the  Kraus operators
\begin{eqnarray}
\mathcal{E}_{ij} = \mathcal{E}_i \otimes \mathcal{E}_j, (i,j=\lbrace 0,1 \rbrace)
\label{A1}
\end{eqnarray}
When finite memory is introduced between two consecutive uses of the amplitude damping channel, the Kraus operators which reproduces Eq.(\ref{mem}) corresponding to the evolution of two-qubits can be constructed as\citep{Yeo}
\begin{eqnarray}
\mathcal{E}_{00} = \begin{pmatrix}
\sqrt{1-p} & 0 & 0 & 0\\
0 & 1 & 0 & 0\\
0 & 0 & 1 & 0\\
0 & 0 & 0 & 1
\end{pmatrix}; ~~
\mathcal{E}_{11} = \begin{pmatrix}
0 & 0 & 0 & 0\\
0 & 0 & 0 & 0\\
0 & 0 & 0 & 0\\
\sqrt{p} & 0 & 0 & 0
\end{pmatrix}
\label{A2}
\end{eqnarray}
which, unlike other decoherence channels can not be decomposed into tensor product of two $2\times 2$ matrices.

\subsection{Phase damping channel with memory}

In this case there is no dissipation of energy from the two-level quantum system. Rather, the environment disperses  keeping the state of the system unchanged modulo damping in phase of the system. Let us denote the $2 \times 2$ identity matrix $\openone_2$ by $\sigma_0$, and the Pauli spinor along the $z$-direction by $\sigma_3$. Now, a memoryless phase-damping channel acting upon the two-qubits has the Kraus operators
\begin{eqnarray}
\mathcal{E}_{ij} = \sqrt{\mathcal{P}_i \mathcal{P}_j} \sigma_i \otimes \sigma_j, (i,j=\lbrace 0,3 \rbrace)
\label{P1}
\end{eqnarray}
where, single qubit Kraus operators are given by, $\mathcal{E}_i = \sqrt{\mathcal{P}_i} \sigma_i$ ($i \in \lbrace 0,3 \rbrace$). And the time-correlated phase-damping channel between two consecutive uses is represented by the Kraus operators\citep{Arrigo}
\begin{eqnarray}
\mathcal{E}_{kk} = \sqrt{\mathcal{P}_k} \sigma_k \otimes \sigma_k, (k=\lbrace 0,3 \rbrace)
\label{P2}
\end{eqnarray} 
where, $\mathcal{P}_0=1-p$ and $\mathcal{P}_3=p$ in which phase-damping parameter $p$ lies within $0\leq p \leq 1$.

\subsection{Depolarizing channel with memory}

In this case the initial state of the system is  decohered due to the presence of errors like bit-flipping, phase-flipping or both. The Kraus operators for a single qubit can be written as
\begin{eqnarray}
\mathcal{E}_i = \sqrt{\mathcal{P}_i} \sigma_i, (i=\lbrace 0,1,2,3 \rbrace)
\end{eqnarray}
where, $\mathcal{P}_0= 1-p$ and $\mathcal{P}_1=\mathcal{P}_2=\mathcal{P}_3=\frac{p}{3}$ with $p$ satisfying $0\leq p\leq 1$, with  $\sigma_0 = \openone_2$,  and $\lbrace \sigma_i \rbrace_{i=1}^3$ are $2\times 2$ Pauli matrices along $x,y$ and $z$ directions, respectively. When the channel has no memory, the evolution of two-qubit is governed by the Kraus operators
\begin{eqnarray}
\mathcal{E}_{ij} = \sqrt{\mathcal{P}_i \mathcal{P}_j} \sigma_i \otimes \sigma_j, (i,j=\lbrace 0,1,2,3 \rbrace)
\label{D1}
\end{eqnarray}
The use of memory in depolarizing channel leads  to the Kraus operators\citep{Macchiavello,Palma}
\begin{eqnarray}
\mathcal{E}_{kk} = \sqrt{\mathcal{P}_k} \sigma_k \otimes \sigma_k, (k=\lbrace 0,1,2,3 \rbrace)
\label{D2}
\end{eqnarray}

\section{Leggett-Garg Inequality in the presence of quantum memory channels} \label{4}

The violation of LGI under decoherence for single qubit system has been studied in Ref\citep{Emary}. Here we consider a two-qubit initial state for which LGI is tested by virtue of measurements performed in subsequent time intervals (i.e. $t_1$,$t_2$,$t_3$ and $t_4$). Generally, a pure bipartite state for qubits has Schmidt-decomposition of the form: 
\begin{eqnarray}
\vert \psi \rangle = k_1 \vert 00 \rangle + k_2 \vert 11 \rangle,
\label{initial}
\end{eqnarray}
where $\vert 0 \rangle, \vert 1 \rangle$ form $\sigma_z$-eigenbasis and $k_1$ and $k_2$ are two non-negative real numbers satisfying $k_1^2 + k_2^2 = 1$, called as Schmidt co-efficients. Hence, we start with the density matrix, $\rho= \vert \psi \rangle\langle \psi \vert$.

\begin{figure}[htp]
\subfigure[temporal correlator $C_{12}$]{\includegraphics[width=\columnwidth, height=3cm,keepaspectratio]{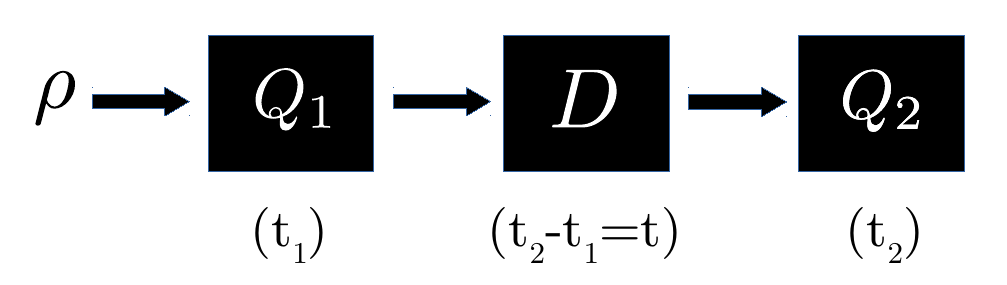}}\\
\subfigure[temporal correlator $C_{23}$]{\includegraphics[width=\columnwidth, height=2.5cm,keepaspectratio]{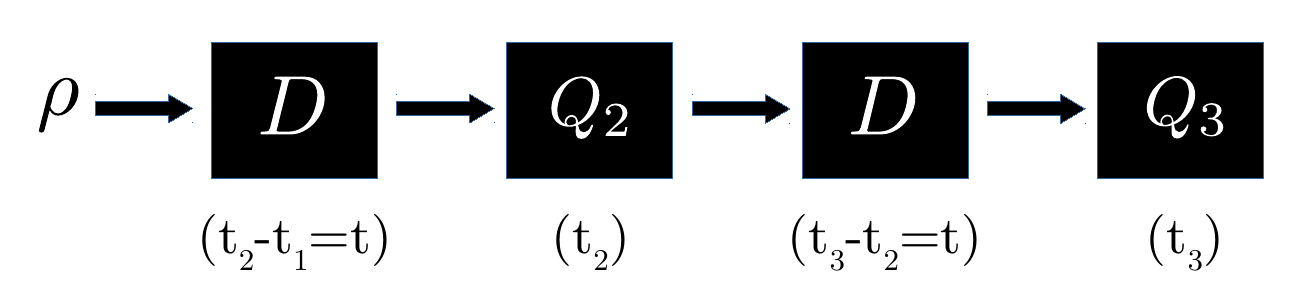}}\\
\subfigure[temporal correlator $C_{34}$]{\includegraphics[width=\columnwidth, height=2cm,keepaspectratio]{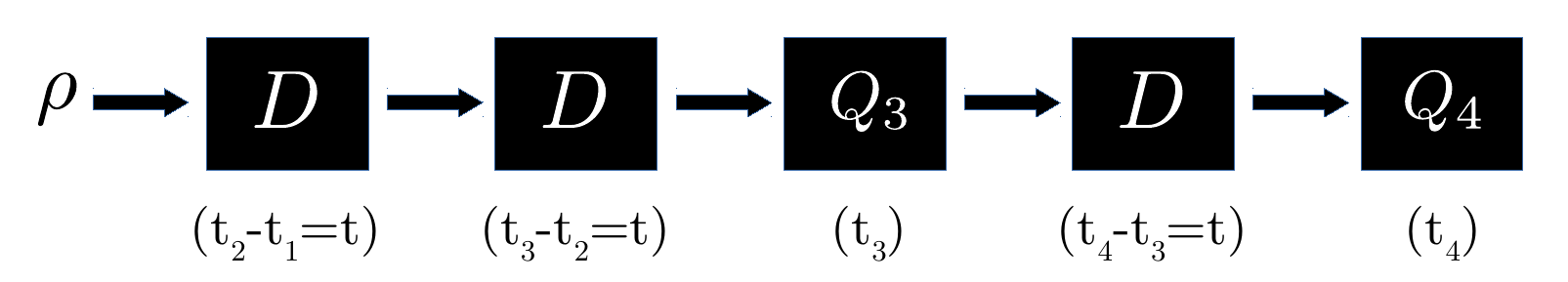}}\\
\subfigure[temporal correlator $C_{14}$]{\includegraphics[width=\columnwidth, height=2cm,keepaspectratio]{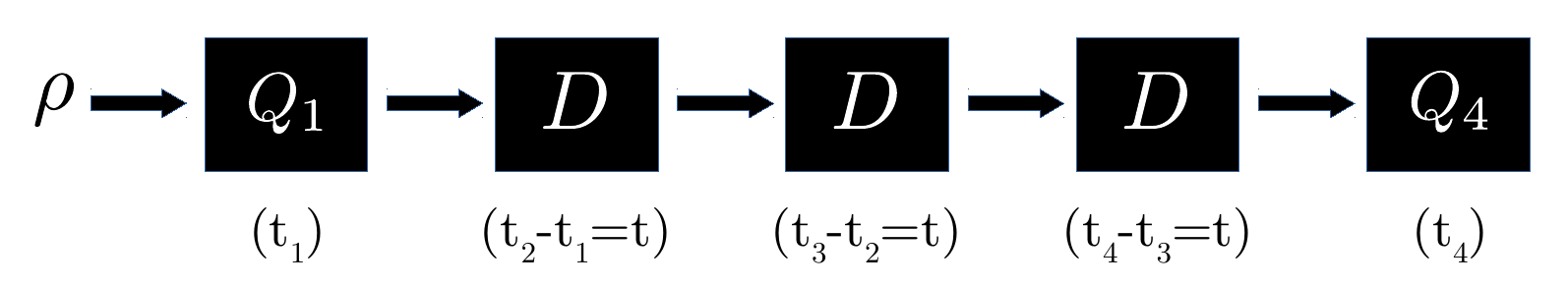}}
\caption{\footnotesize Schematic diagram for evaluation of 4-term LGI using the initial state $\rho$, Bell-state measurements $Q_i$ at time $t_i$, and quantum memory channels $D$ at subsequent intervals $t$.}
\label{Depict}
\end{figure}

To calculate the 4-term LGI given by Eq.(\ref{LGI}), we consider all the measurement time intervals equal. There are time evolutions of the composite system between two measurements. Here, the evolution is governed by the decoherence having  parameter $p$. The mechanism is pictorially represented in Fig.\ref{Depict}.

Following \citep{lobejko} we consider a very natural first choice experiment that can be performed on a two-qubit state, {\it i.e.}, the Bell State Measurement(BSM)\citep{Pan}. It is to be noted that here we consider BSM in generalized directions. The projectors corresponding to four generalised Bell states can be written as
\begin{align}
& P_{\psi^+} = \vert \psi^+ \rangle\langle \psi^+ \vert \nonumber\\
& P_{\psi^-} = \vert \psi^- \rangle\langle \psi^- \vert \nonumber\\
& P_{\phi^+} = \vert \phi^+ \rangle\langle \phi^+ \vert \nonumber\\
& P_{\phi^-} = \vert \phi^- \rangle\langle \phi^- \vert 
\end{align} 
where, $\vert\psi^+\rangle=\frac{1}{\sqrt{2}}(\vert0_n 1_n\rangle+\vert1_n 0_n\rangle)$, $\vert\psi^-\rangle=\frac{1}{\sqrt{2}}(\vert0_n 1_n\rangle-\vert1_n 0_n\rangle)$, $\vert\phi^+\rangle=\frac{1}{\sqrt{2}}(\vert0_n 0_n\rangle+\vert1_n 1_n\rangle)$ and $\vert\phi^-\rangle=\frac{1}{\sqrt{2}}(\vert0_n 0_n\rangle-\vert1_n 1_n\rangle)$. The unit vectors $\vert0_n\rangle=\cos(\frac{\theta}{2})\vert 0 \rangle+e^{i \phi}\sin(\frac{\theta}{2})\vert 1 \rangle$ and $\vert1_n\rangle=-e^{-i \phi}\sin(\frac{\theta}{2})\vert 0 \rangle+\cos(\frac{\theta}{2})\vert 1 \rangle$ are eigenstates of $\widehat{n}.\overline{\sigma}$ where $\widehat{n}\equiv (\sin\theta \cos\phi, \sin\theta \sin\phi, \cos\theta)$, and $\overline{\sigma}\equiv (\sigma_x,\sigma_y,\sigma_z)$ is the Pauli vector. In this case there will be more than one measurement scheme and assignment of value $\{1,-1\}$ to outcomes, which we specify as follows.

\begin{enumerate}
\item {\it Type-I:} The usual modifications of BSM having restricted scope of distinguishability among four Bell states consists of measurements corresponding to four one-dimensional projectors in non-degenerate subspaces. We assign value +1 if either one of the projectors $P_{\psi^+}$ or $P_{\phi^+}$ clicks and we set value -1 for either one of the projectors $P_{\psi^-}$ or $P_{\phi^-}$.
\item {\it Type-II:} Here we consider two projectors corresponding to outcomes +1 and -1, which are
\begin{align}
& P_+ = \vert \psi^+ \rangle\langle \psi^+ \vert, \nonumber\\
& P_- = \openone_4 - P_+
\end{align}
Here, $P_-$ belongs to the $4$-dimensional degenerate subspace spanned by three Bell-states, and only one of the four generalised Bell states can be distinguished by this way.
\item {\it Type-III:} A less precise degenerate BSM consists of two $2$-dimensional projectors which are
\begin{align}
& P_+ = \vert \psi^+ \rangle\langle \psi^+ \vert + \vert \phi^+ \rangle\langle \phi^+ \vert, \nonumber\\
& P_- = \openone_4 - P_+
\end{align}
corresponding to outcomes +1 and -1 respectively. By performing such measurement, the only information we can extract is in which of the subspaces spanned by a pair of generalised Bell states the measured system lies. Though the value assignment here is similar to that of Type-I measurements, the degeneracy of the projector subspace is the key difference between the two.
\end{enumerate} 

To analyse the LG test, we assume $Q_1,Q_2,Q_3$ and $Q_4$ as BSMs in four different directions $\widehat{n}_1\equiv (\sin\theta_1 \cos\phi_1, \sin\theta_1 \sin\phi_1, \cos\theta_1)$, $\widehat{n}_2\equiv (\sin\theta_2 \cos\phi_2, \sin\theta_2 \sin\phi_2, \cos\theta_2)$, $\widehat{n}_3\equiv (\sin\theta_3 \cos\phi_3, \sin\theta_3 \sin\phi_3, \cos\theta_3)$ and $\widehat{n}_4\equiv (\sin\theta_4 \cos\phi_4, \sin\theta_4 \sin\phi_4, \cos\theta_4)$. The choices of $\theta_i$ and $\phi_i$ are arbitrary within the constraints $0 \leq \theta_i \leq \pi$ and $0 \leq \phi_i \leq 2\pi$ (i=1,2,3,4). We, now  illustrate the action of quantum decoherence channels, i.e.,  amplitude damping channel, phase-damping channel and depolarising channel in a LGI test applied on the initial state given by Eq.(\ref{initial}).

\subsection{Amplitude damping channel with memory}

In order to calculate the four-term LGI $K_4$, we choose the decoherence parameter $p$ to be the same for every use of the amplitude damping channel between two consecutive measurements. Following the different measurement set ups, we evaluate the QM maximum of LGI and at the optimal points of $\theta_i$s and $\phi_i$s (i=1,2,3,4), we obtain the dynamics of $K_4$ with respect to $p$ and $\mu$ given below.

We first calculate  $K_4$ according to {\it Type-I} generalised Bell-State Measurements and Kraus operators given by Eq.(\ref{A1}) and Eq.(\ref{A2}). We find that the QM maximum of $K_4$ is $3.18$  occurs at the values of measurement parameters, $\theta_1=1.88$, $\phi_1=0.77$, $\theta_2=1.54$,$\phi_2=0.57$, $\theta_3=1.21$, $\phi_3=0.21$, $\theta_4=3.14$, $\phi_4=1.73$ (all are in Radians) and when the initial state is $\frac{1}{\sqrt{2}}(\vert 00 \rangle + \vert 11 \rangle)$. It may be mentioned here that it is possible for the maximum violation of the LGI to exceed the bound of $2\sqrt{2}$ obtained in the case of non-degenerate projective measurements on a qubit, as noted in earlier works \citep{b2, Dakic}. Using these parameter values, we can write the function as
\begin{align}
& K_4 = \nonumber\\
& 2.27 + 0.91 \sqrt{1-p} + \mu [-3.23 + 3.23 \sqrt{1-p} \nonumber\\
& + \mu \lbrace7.18 - 7.18 \sqrt{1-p} + (-4.84 + 4.84 \sqrt{1-p}) \mu\rbrace] \nonumber\\
& + p^5 [0.07 + \mu \lbrace-0.22 + (0.22 - 0.07 \mu) \mu\rbrace] \nonumber\\
& + p \big(-5.38 - 1.02 \sqrt{1-p} + \mu [12.41 - 6.12 \sqrt{1-p} \nonumber\\
& + \mu \lbrace-18.04 + 14.44 \sqrt{1-p} + (10.94 - 8.52 \sqrt{1-p}) \mu\rbrace]\big) \nonumber\\
& + p^3 \big(-1.64 - 0.01 \sqrt{1-p} + \mu [4.86 - 0.30 \sqrt{1-p} \nonumber\\
& + \mu \lbrace-4.64 + 0.55 \sqrt{1-p} + (1.42 - 0.24 \sqrt{1-p}) \mu\rbrace]\big) \nonumber\\
& + p^4 \big(0.82 + \mu [-1.51 + 0.08 \sqrt{1-p} \nonumber\\
& + \mu \lbrace0.54 - 0.16 \sqrt{1-p} + (0.14 + 0.08 \sqrt{1-p}) \mu\rbrace]\big) \nonumber\\
& + p^2 \big(4.55 + 0.12 \sqrt{1-p} + \mu [-11.80 + 3.66 \sqrt{1-p} \nonumber\\
& + \mu \lbrace14.74 - 7.62 \sqrt{1-p} + (-7.49 + 3.84 \sqrt{1-p}) \mu\rbrace]\big)
\end{align}

\begin{figure}[t]
\centering
\subfigure[Type-I BSM]{
\includegraphics[width=.225\textwidth]{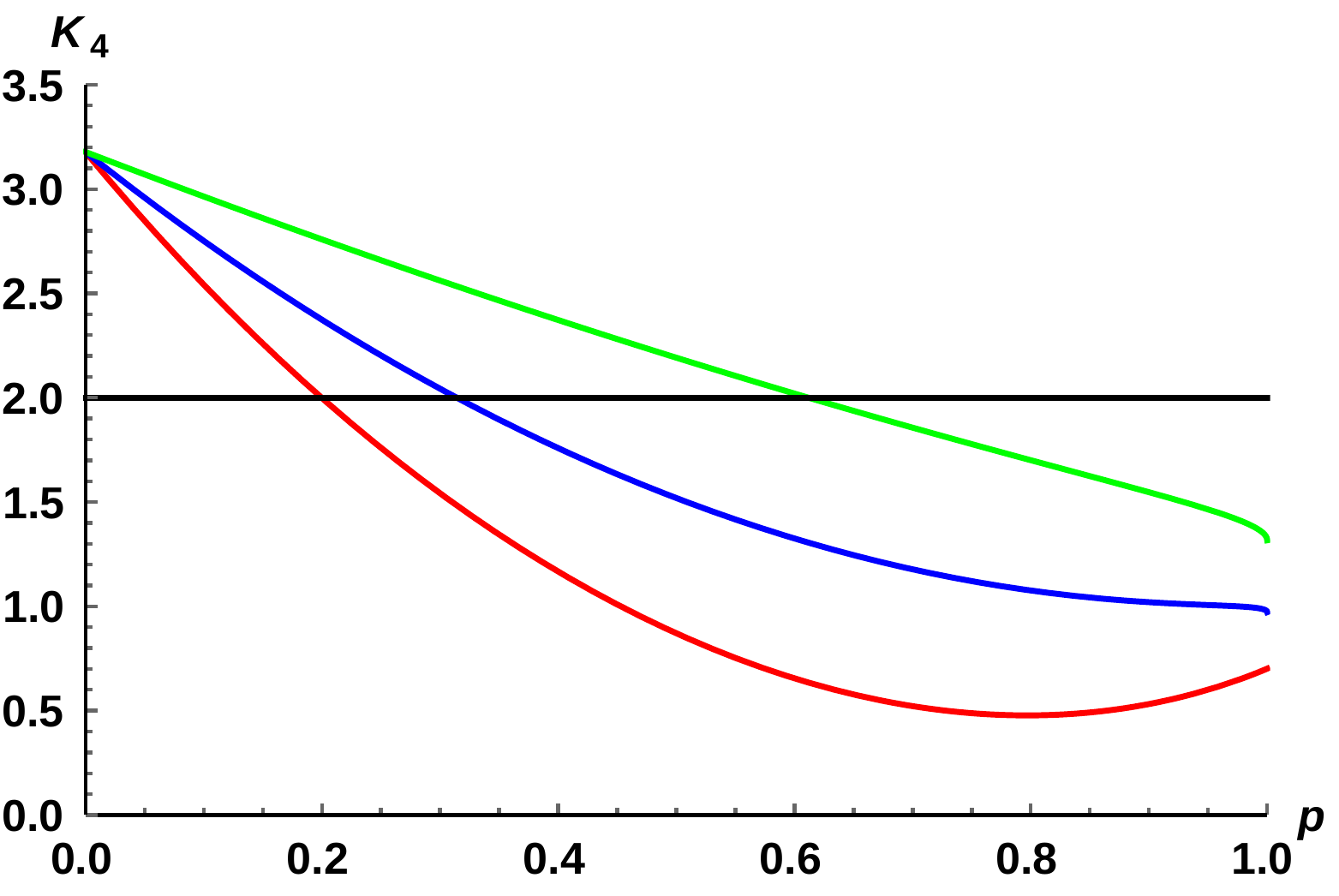}
\label{AL1}
}
\subfigure[Type-II BSM]{
\includegraphics[width=.225\textwidth]{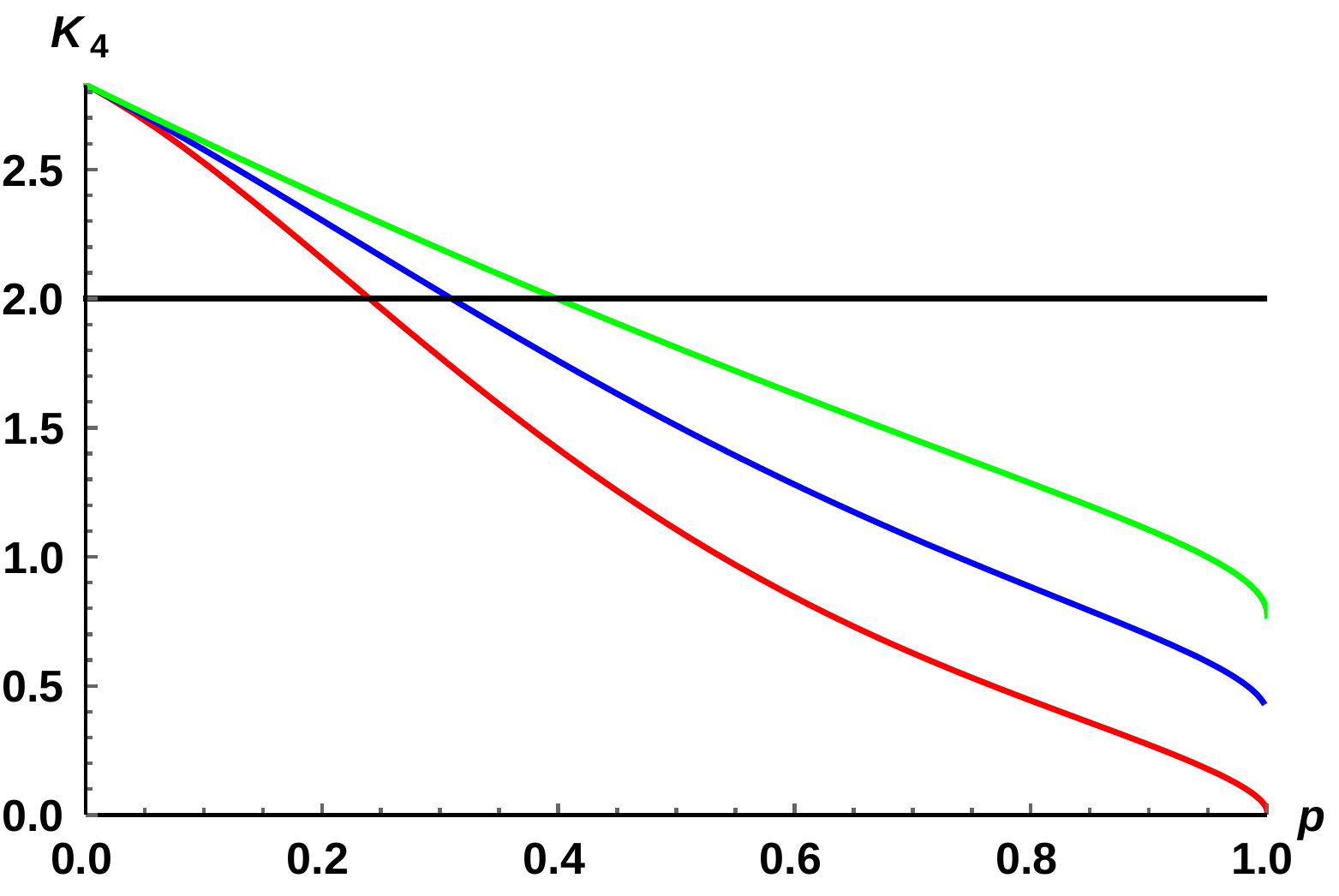}
\label{AL2}
}
\subfigure[Type-III BSM]{
\includegraphics[width=.225\textwidth]{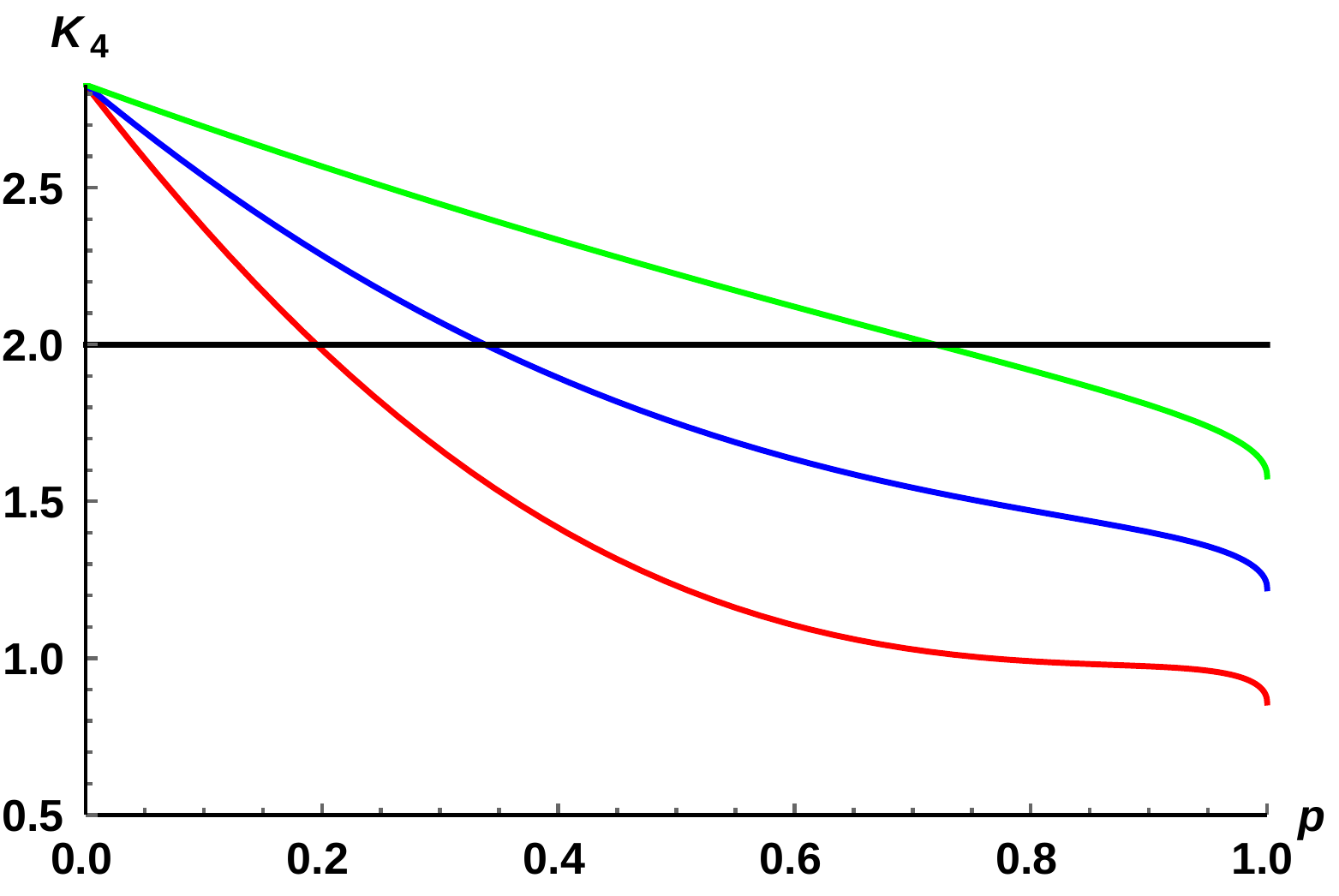}
\label{AL3}
}
\caption{\footnotesize (Coloronline) Plot of $K_4$ vs amplitude damping strength $p$ for memory co-efficients $\mu=0$ (bottom curve), $\mu=\frac{1}{2}$ (intermediate curve) and $\mu=1$ (top curve) using three types of generalised Bell State Measurements. The horizontal line indicates the classical bound.}
\label{ADC-LGI}
\end{figure}

We plot $K_4$ w.r.t. the damping parameter $p$ for three different values of the memory co-efficient,$\mu$ in Fig.\ref{AL1}. We see that the behaviour of $K_4$ is always non-increasing with increasing decoherence, but if memory is introduced to the amplitude-damping channel, then one can preserve the violation of LGI against decoherence with increasing magnitude for $\mu$ from zero (no memory) to $1$ (perfect memory). 

Similar to the above, we calculate $K_4$ using the {\it Type-II} generalised Bell State Measurement, and its QM maximum is $2\sqrt{2}$ which occurs at the  measurement angles, $\theta_1=1.86$,$\phi_1=1.23$, $\theta_2=1.01$, $\phi_2=4.06$, $\theta_3=2.38$, $\phi_3=0.51$, $\theta_4=1.38$,$\phi_4=3.08$  and for Schmidt co-efficients, $k_1=\frac{1}{\sqrt{2}}$ and $k_2=\frac{1}{\sqrt{2}}$ of the initial state. At these values, $K_4$ turns out to be
\begin{align}
& K_4= \nonumber\\
& 2.11 + 0.72 \sqrt{1-p} + \mu [-4.55 + 4.55 \sqrt{1-p} + \mu \lbrace7.89 \nonumber\\
& - 7.89 \sqrt{1-p} + (-4.99 + 4.99 \sqrt{1-p}) \mu\rbrace] + p^6 [-1.07 \nonumber\\
& + \mu \lbrace3.22 + \mu (-3.22 + 1.07 \mu)\rbrace] + p \big(-1.67  \nonumber\\
& - 0.36 \sqrt{1-p} + \mu [10.67 - 8.25 \sqrt{1-p} + \mu \lbrace-19.58 \nonumber\\
& + 15.63 \sqrt{1-p} + (10.98 - 8.48 \sqrt{1-p}) \mu\rbrace]\big) \nonumber\\
& + p^3 \big(17.82 - 0.04 \sqrt{1-p} + \mu [-32.80 - 0.69 \sqrt{1-p} \nonumber\\
& + \mu \lbrace16.70 + 1.07 \sqrt{1-p} + (-1.72 - 0.34 \sqrt{1-p}) \mu\rbrace]\big) \nonumber\\
& + p^5 [6.22 + \mu \lbrace-16.52 + \mu (14.37 - 4.08 \mu)\rbrace] \nonumber\\
& + p^4 \big(-15.24 + 0.01 \sqrt{1-p} + \mu [35.23 + 0.31 \sqrt{1-p} \nonumber\\
& + \mu \lbrace-25.83 - 0.66 \sqrt{1-p} + (5.83 + 0.34 \sqrt{1-p}) \mu\rbrace]\big) \nonumber\\
& + p^2 \big(-8.16 + 0.23 \sqrt{1-p} + \mu [5.49 + 4.55 \sqrt{1-p} \nonumber\\
& + \mu \lbrace 9.67 - 8.17 \sqrt{1-p} + (-7.08 + 3.47 \sqrt{1-p}) \mu\rbrace]\big)
\end{align}

In Fig.\ref{AL2}, we observe that $K_4$ decreases with the dissipation strength $p$. However, by increasing memory strength $\mu$ from 0 to 1, it can be seen that, the violation of LGI (Eq.\ref{LGI}) increases with $\mu$ and thereby the diminishing effect of decoherence is checked to some extent.

Following the same procedure, the left hand side of Eq.(\ref{LGI}) can be calculated while using generalised Bell State Measurement of {\it Type-III}. In this case too, we find the optimal QM value of $K_4$ to be $2\sqrt{2}$. This achieved for the measurement parameters, $\theta_1=0.48$, $\phi_1=0.34$,$\theta_2=1.44$, $\phi_2=0.60$, $\theta_3=2.04$, $\phi_3=0.79$, $\theta_4=2.63$, $\phi_4=0.82$ and the state variables, $k_1=\frac{1}{\sqrt{2}}$ and $k_2=\frac{1}{\sqrt{2}}$. Using these values, we can write the function $K_4$ as

\begin{align}
& K_4= \nonumber\\
& 1.65 + 1.17 \sqrt{1-p} + \mu [-0.16 + 0.16 \sqrt{1-p} + \mu \lbrace1.11 \nonumber\\
& - 1.11 \sqrt{1-p} + (-1.50 + 1.50 \sqrt{1-p}) \mu\rbrace] + p^6 [-0.13 \nonumber\\
& + \mu \lbrace 0.40 + (-0.40 + 0.13 \mu) \mu\rbrace] + p \big(-2.44 \nonumber\\
& - 1.84 \sqrt{1-p} + \mu [3.89 - 0.31 \sqrt{1-p} + \mu \lbrace -3.69 \nonumber\\
& + 3.14 \sqrt{1-p} + (2.88 - 2.13 \sqrt{1-p}) \mu\rbrace]\big) + p^3 \big(3.10 \nonumber\\
& - 0.49 \sqrt{1-p} + \mu [-5.30 - 1.11 \sqrt{1-p} + \mu \lbrace 2.19 \nonumber\\
& + 2.76 \sqrt{1-p} + (0.01 - 1.16 \sqrt{1-p}) \mu\rbrace]\big) + p^5 [1.04 \nonumber\\
& + \mu \lbrace -2.84 + \mu (2.57 + -0.77 \mu)\rbrace] + p^4 \big(-2.66 \nonumber\\
& - 0.02 \sqrt{1-p} + \mu [6.30 + 0.84 \sqrt{1-p} + \mu \lbrace-4.76 \nonumber\\
& - 1.62 \sqrt{1-p} + (1.12 + 0.80 \sqrt{1-p}) \mu\rbrace]\big) + p^2 \big(0.31 \nonumber\\
& + 1.89 \sqrt{1-p} + \mu [-1.56 + 0.35 \sqrt{1-p} + \mu \lbrace 2.98 \nonumber\\
& - 3.09 \sqrt{1-p} + (-1.88 + \sqrt{1-p}) \mu \rbrace]\big)
\end{align}

It is depicted in Fig.\ref{AL3} that as the damping strength $p$ increases, the violation of LGI decreases. The use of memory with strength $\mu$ ranging from 0 to 1, counters the  effect of decoherence, and thereby protects the temporal correlation from falling below the classical bound
up to a large value of dissipation.

\subsection{Phase damping channel with memory}

We consider that the composite system $\rho$ passes through the environment with the same Phase-damping parameter $p$ and the memory parameter $\mu$ during any time interval $t_{k+1}-t_k$ (k=1,2,3). Using the Kraus operators given by Eqs.(\ref{P1}, \ref{P2}) and three types of measurement schemes, we are able to determine the quantum mechanical optimum of $K_4$ and considering the measurements and the state which makes it optimal, we can figure out $K_4$ as a function of $p$ and $\mu$.

\begin{figure}[t]
\centering
\subfigure[Type-I BSM]{
\includegraphics[width=.225\textwidth]{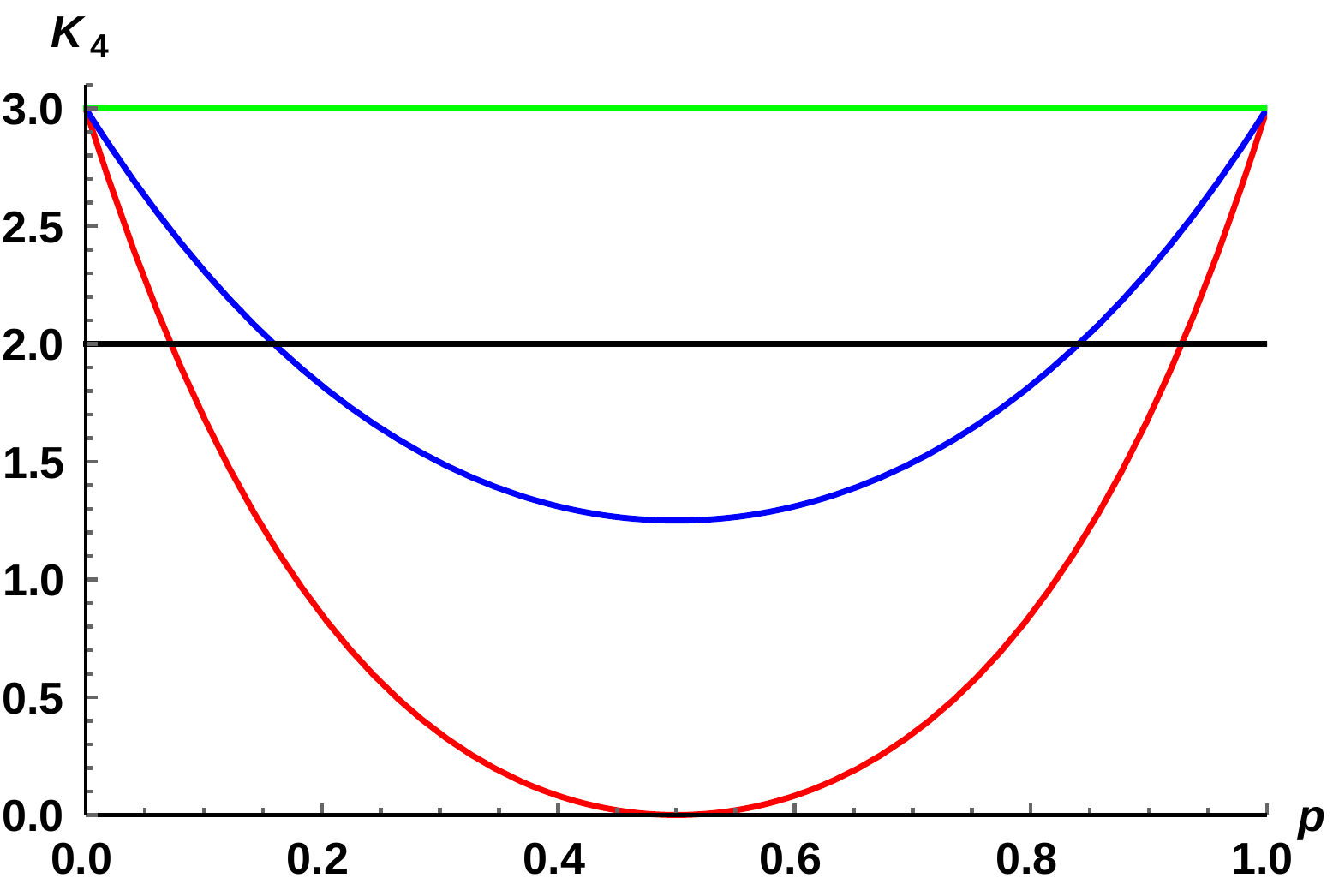}
\label{PL1}
}
\subfigure[Type-II BSM]{
\includegraphics[width=.225\textwidth]{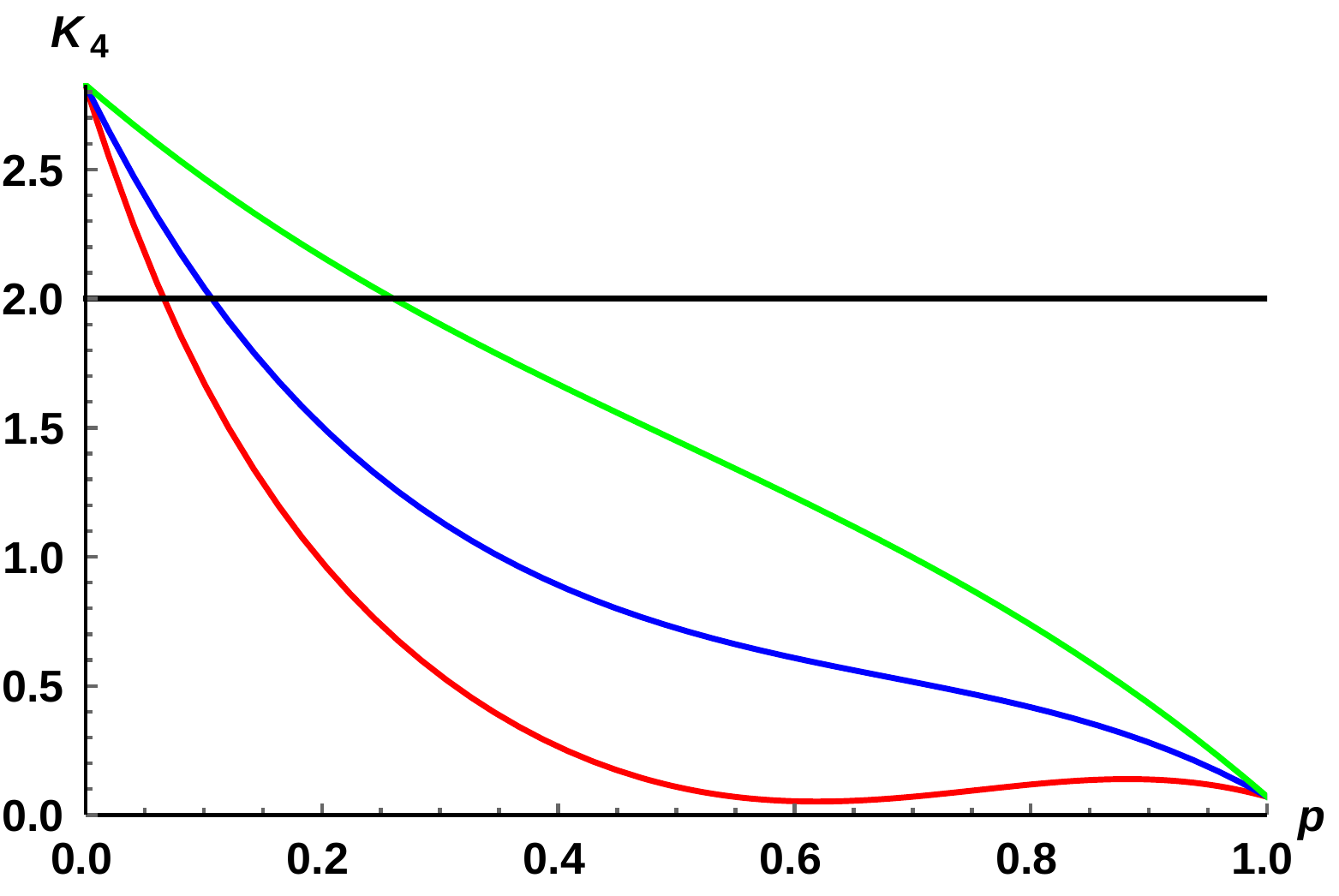}
\label{PL2}
}
\subfigure[Type-III BSM]{
\includegraphics[width=.225\textwidth]{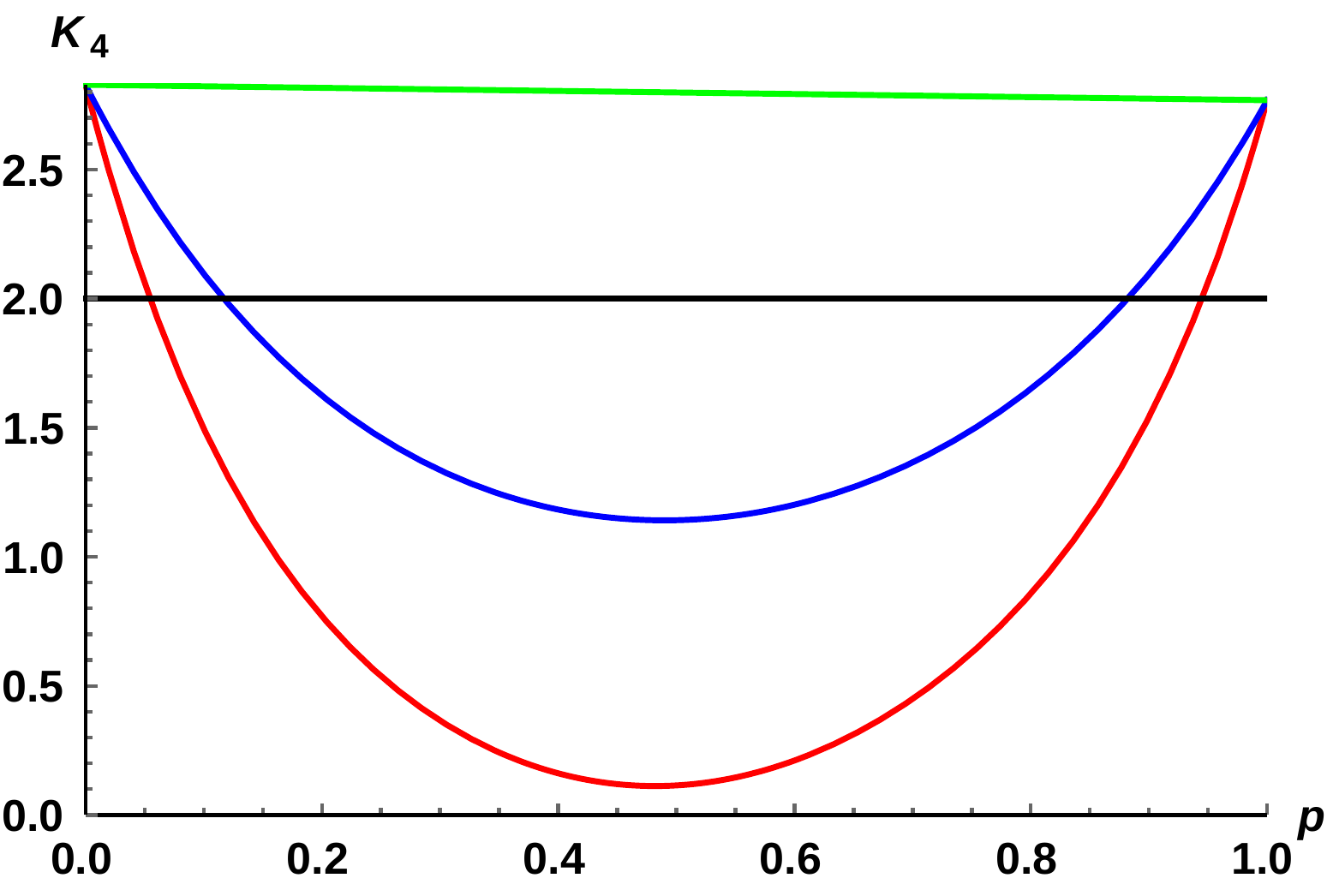}
\label{PL3}
}
\caption{\footnotesize (Coloronline) $K_4$ is plotted against the phase-damping co-efficient $p$ at memory strengths, $\mu=0$(bottom red curve), $\mu=\frac{1}{2}$(middle blue curve) and $\mu=1$(top green curve) applying generalised Bell State Measurements of {\it Type-I, II} and {\it III}}
\label{PDC-LGI}
\end{figure}

We first focus on {\it Type-I} BSM, for which the maximum QM violation of the LGI is found to be $3$. One can reach this amount of violation by setting the measurement angles, $\theta_1=2.36$, $\phi_1=1.57$,$\theta_2=0.01$, $\phi_2=1.66$, $\theta_3=0$, $\phi_3=1.31$, $\theta_4=1.57$, $\phi_4=0$ and the Schmidt co-efficients of initial state, $k_1=\frac{1}{\sqrt{2}}$ and $k_2=\frac{1}{\sqrt{2}}$. For the above value values, $K_4$ can be expressed as
\begin{align}
& K_4= \nonumber\\
& 3 + p \big(-16 + 16 \mu + p [32 + \mu (-48 + 16 \mu) \nonumber\\
& + p \lbrace-32 + \mu (64 -32 \mu)\rbrace + p^2 \lbrace 16 + \mu (-32 + 16 \mu)\rbrace]\big)
\label{constviol}
\end{align}
We show the behaviour of $K_4$ with the phase-damping co-efficient $p$ in  Fig.\ref{PL1}. Though it is not always decreasing with $p$, we see that by increasing the  memory strength $\mu$, one can preserve the violation of LGI for a larger range of phase damping strength. In fact, for $\mu=1$, the magnitude of violation stays at its maximum independent of the decoherence strength $p$, as is evident from the Eq.(\ref{constviol}).

Now let us study the case of {\it Type-II} BSM. In this scenario, the maximum value of $K_4$ is  $2\sqrt{2}$. It happens for $\theta_1=1.27$,$\phi_1=0.63$,$\theta_2=1.22$,$\phi_2=3.85$,$\theta_3=2.28$,$\phi_3=1.22$,$\theta_4=2.16$,$\phi_4=1.84$ and $k_1=k_2=\frac{1}{\sqrt{2}}$. Thus, the function becomes
\begin{align}
& K_4= \nonumber\\
& 2.83 + 17.65 p^6 (1 - \mu)^3 + p^4 (99.63 - 252.22 \mu \nonumber\\
& + 205.56 \mu^2 - 52.96 \mu^3) + p^2 (40.17 - 54.07 \mu \nonumber\\
& + 17.47 \mu^2) + p (-14.86 + 10.91 \mu) + p^3 (-77.79 \nonumber\\
& + 160.26 \mu - 102.51 \mu^2 + 17.65 \mu^3) + p^5 (-67.56 \nonumber\\
& + 188.08 \mu - 173.49 \mu^2 + 52.96 \mu^3)
\end{align}

In Fig.\ref{PL2}, it is clearly seen that $K_4$ decreases with increasing strength of phase damping $p$. As the memory co-efficient $\mu$ increases, the violation of LGI  improves.

The use of {\it Type-III} measurement leads to a similar character of the function $K_4$ as in the case of {\it Type-I} measurement. It has QM maximum of the value $2\sqrt{2}$, which is achieved by the measurement angles, $\theta_1=2.76$, $\phi_1=1.29$, $\theta_2=0.81 $ ,$\phi_2=0.87$, $\theta_3=1.86$,$\phi_3=1.78$, $\theta_4=3.14$, $\phi_4=0.32$ and initial state parameters, $k_1=k_2=\frac{1}{\sqrt{2}}$. These makes $K_4$ of the form:
\begin{align}
& K_4 = \nonumber\\
& 2.83 + 45.94 p^6 (1 - \mu)^3 + p^4 (175.19 - 488.19 \mu \nonumber\\
& + 450.82 \mu^2 - 137.82 \mu^3) + p (-17.91 + 17.85 \mu) \nonumber\\
& + p^2 (56.50 - 93.87 \mu + 37.37 \mu^2) + p^3 (-121.96 \nonumber\\
& + 288.57 \mu - 212.55 \mu^2 + 45.94 \mu^3) + p^5 (-137.82 \nonumber\\
& + 413.45 \mu - 413.45 \mu^2 + 137.82 \mu^3)
\end{align}

We see in Fig.\ref{PL3} that $K_4$  first decreases with damping parameter and then increases with $p$. As expected, the violation of LGI increases with the strength of memory, $\mu$, and stays constant for $\mu=1$. Thus, memory channels are able to protect temporal correlations against phase damping.

\subsection{Depolarising channel with memory}

Here we consider the depolarising channel through which the two-qubit state is transmitted. Let us assume that the depolarising strength, $p$ is unaltered between any pair of generalised BSMs. Along with various schemes of BSMs, Kraus operations using Eq.(\ref{D1},\ref{D2}), one can easily obtain the QM optimal value of the left hand side of Eq.(\ref{LGI}). Then using the critical measurement angles and state estimators of this global extremum, the nature of the function $K_4$ can be explored w.r.t. the decoherence strength $p$ and the memory parameter $\mu$.

Choosing first the {\it Type-I} measurement settings to evaluate $K_4$,  the maximum QM violation of Eq.(\ref{LGI}) turns out to be $3.18$, which occurs at the measurement variables, $\theta_1=1.88$,$\phi_1=0.77$, $\theta_2=1.54$, $\phi_2=0.57$, $\theta_3=1.21$, $\phi_3=0.21$ ,$\theta_4=3.14$,$\phi_4=1.73$ and state co-efficients, $k_1=k_2=\frac{1}{\sqrt{2}}$. We find that
\begin{align}
& K_4= \nonumber\\
& 3.18 + 7.27 p^6 (1 - \mu)^3 + p (-15.32 + 12.54 \mu) \nonumber\\
& + p^2 (37.68 - 57.20 \mu + 21.14 \mu^2) \nonumber\\
& + p^4 (61.28 - 158.12 \mu + 135.26 \mu^2 - 38.42 \mu^3) \nonumber\\
& + p^3 (-61.15 + 131.41 \mu - 88.96 \mu^2 + 17.98 \mu^3) \nonumber\\
& + p^5 (-32.72 + 93.65 \mu - 89.14 \mu^2 + 28.21 \mu^3)
\end{align}

\begin{figure}[t]
\centering
\subfigure[Type-I BSM]{
\includegraphics[width=.225\textwidth]{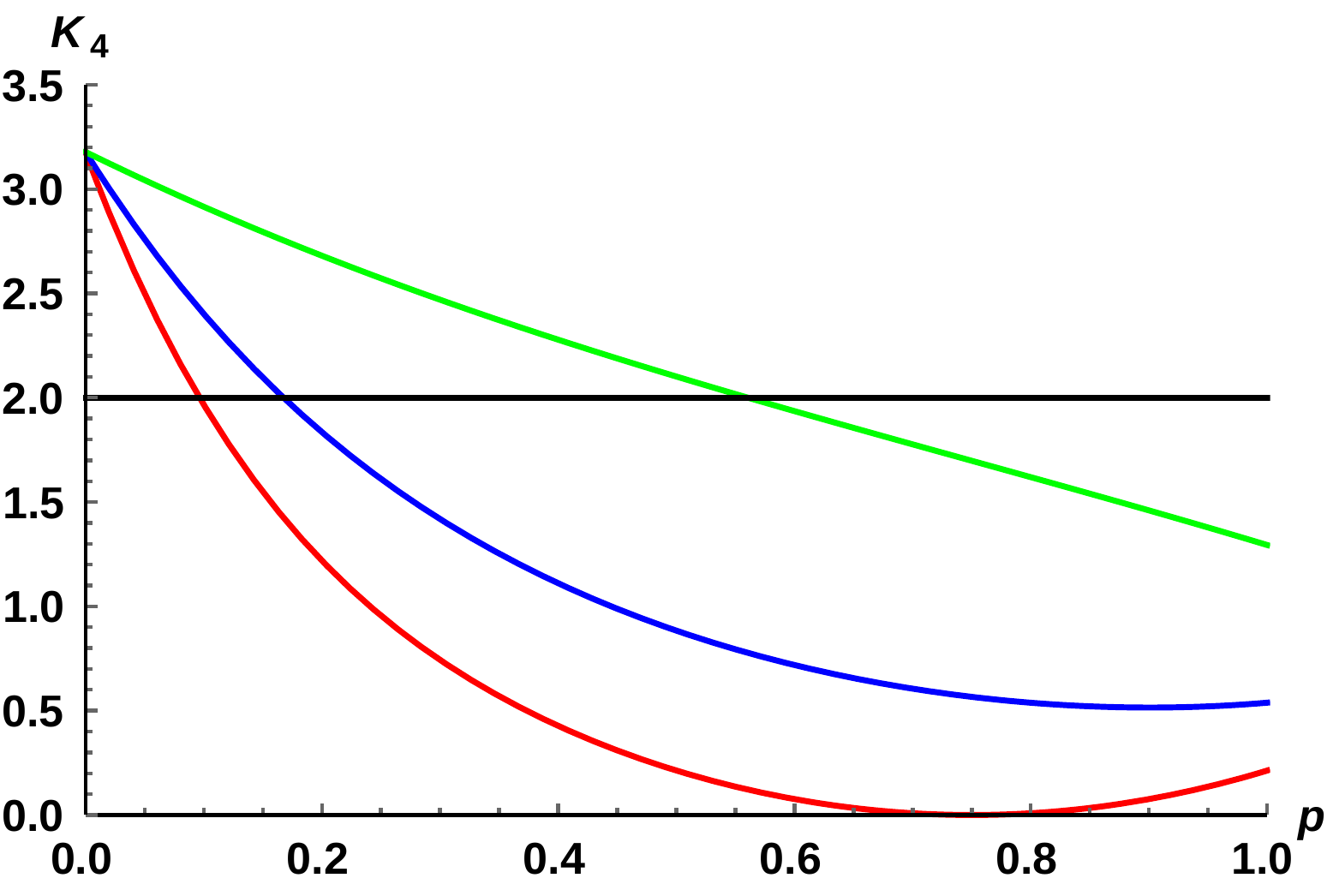}
\label{DL1}
}
\subfigure[Type-II BSM]{
\includegraphics[width=.225\textwidth]{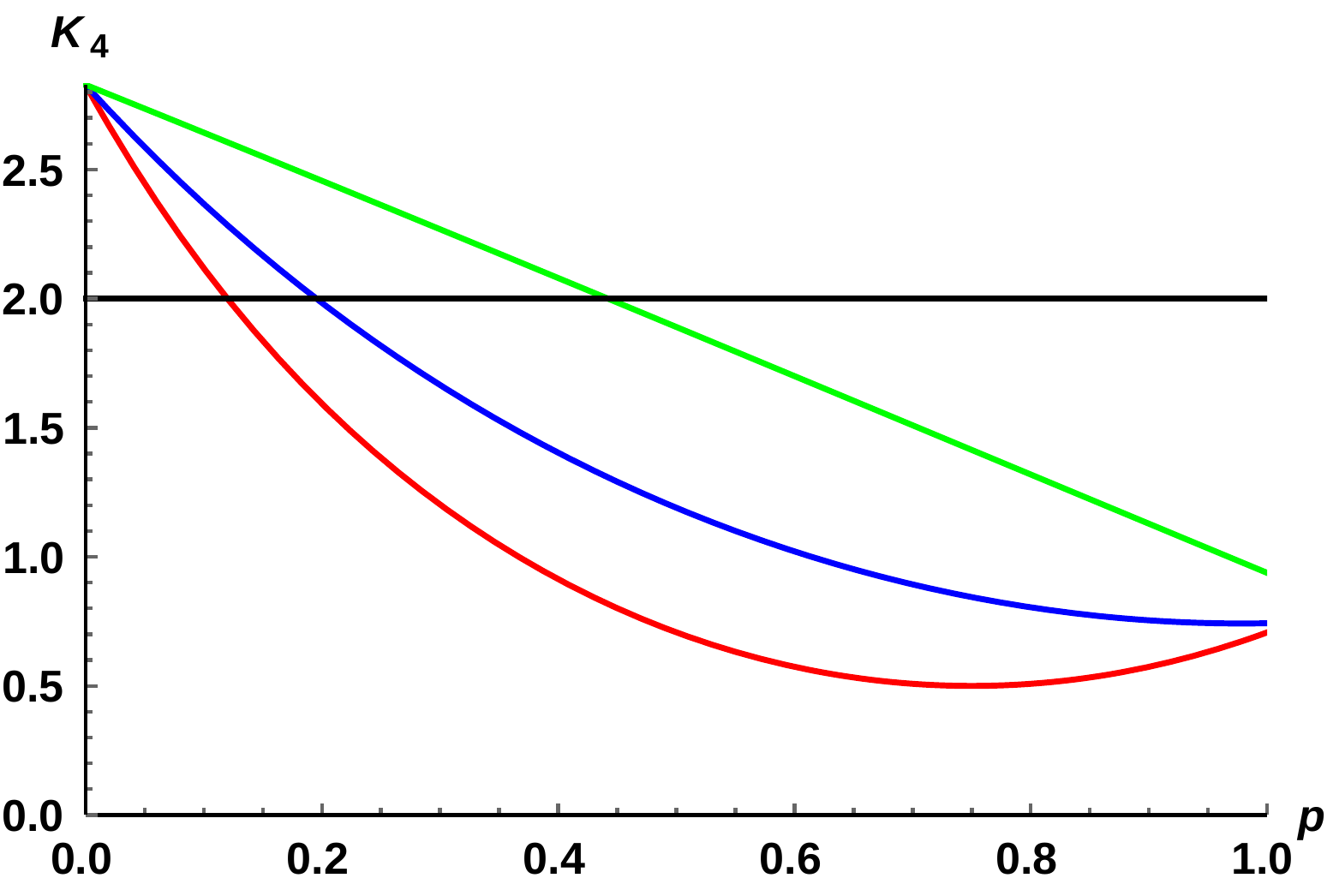}
\label{DL2}
}
\subfigure[Type-III BSM]{
\includegraphics[width=.225\textwidth]{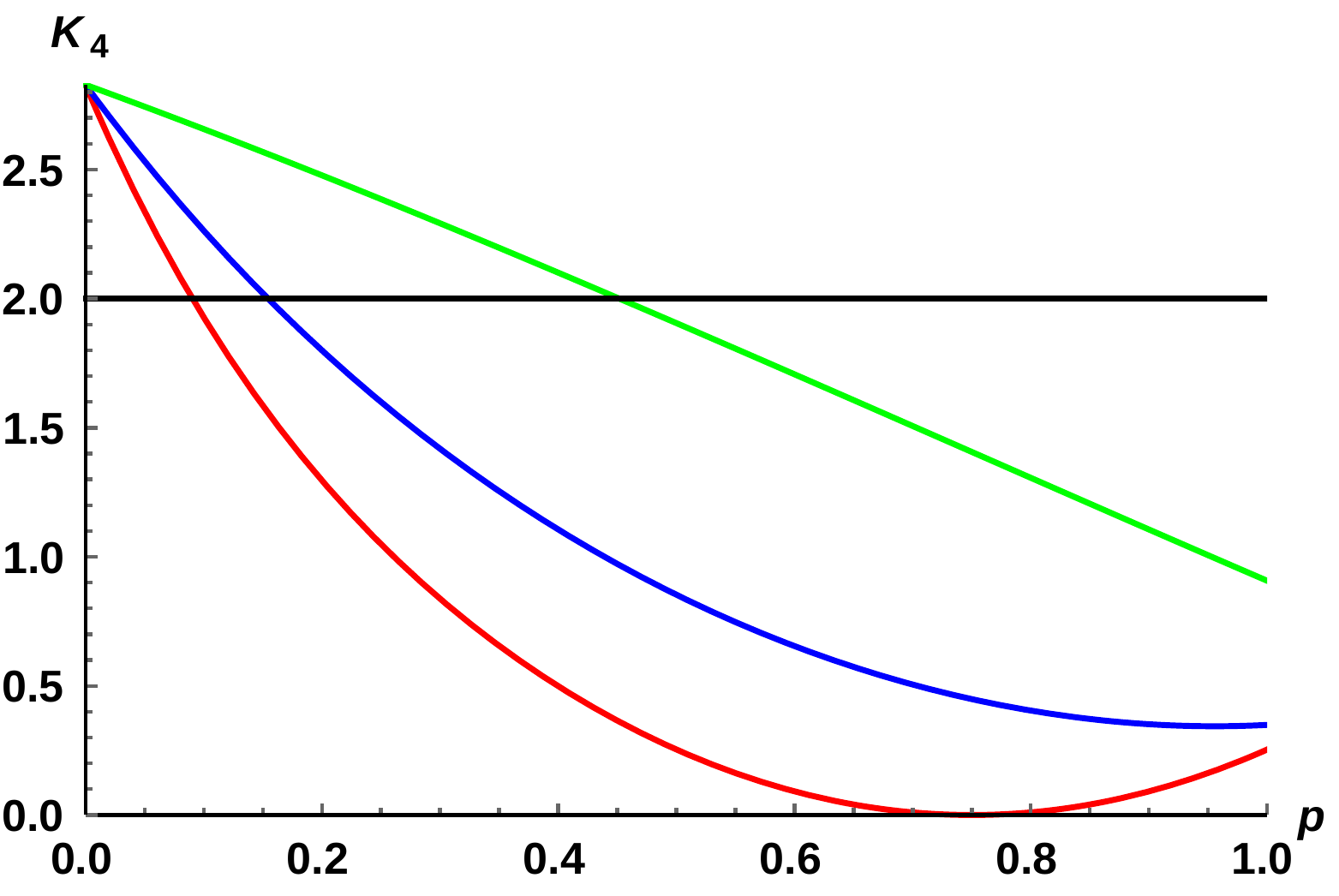}
\label{DL3}
}
\caption{\footnotesize (Coloronline) Plot of $K_4$ vs depolarising parameter, $p$ with memory, $\mu=0$(bottom red curve), $\mu=\frac{1}{2}$(middle blue curve) and $\mu=1$(top green curve) by using three types of generalised Bell State Measurements}
\label{DC-LGI}
\end{figure}

We plot $K_4$ with depolarising strength $p$ in Fig.\ref{DL1} where one can observe that the violation of LGI decreases with $p$ until one reaches near the maximum decoherence strength. However, as memory is increased during environmental interaction, the violation of LGI can be preserved to some extent countering the environmental effect.

Now, we consider the {\it Type-II} measurement scheme and applying this, we get maximum QM value for $K_4$ to be $2\sqrt{2}$ which occurs at the values of measurement parameters, $\theta_1=1.24$,$\phi_1=1.37$, $\theta_2=0.91$, $\phi_2=1.13$, $\theta_3=0.63$, $\phi_3=0.73$, $\theta_4=0.03$,$\phi_4=0.47$ along with Schmidt co-efficients, $k_1=\frac{1}{\sqrt{2}}$ and $k_2=\frac{1}{\sqrt{2}}$ of the initial state, $\rho$. Hence, we obtain 

\begin{align}
& K_4= \nonumber\\
& 2.83 + 1.56 p^6 (1 - \mu)^3 + p (-8.25 + 6.41 \mu) \nonumber\\
& + p^2 (12.94 - 20.03 \mu + 7 \mu^2) + p^4 (13.87 - 39.02 \mu \nonumber\\
& + 36.30 \mu^2 - 11.14 \mu^3) + p^3 (-15.20 + 35.99 \mu \nonumber\\
& - 26.39 \mu^2 + 5.64 \mu^3) + p^5 (-7.04 + 21.37 \mu \nonumber\\
& - 21.61 \mu^2 + 7.28 \mu^3)
\end{align}

We see again in Fig.\ref{DL2} that the environment decreases temporal correlations observed by the of violation of Eq.(\ref{LGI}), and to counterbalance this, increasing amount of memory $\mu$ is effective in protecting such correlations.

In a similar way, we consider {\it Type-III} generalised Bell State Measurements and obtain the QM extremum value of $K_4$ to be $2\sqrt{2}$. It occurs for the angles, $\theta_1=2.76$, $\phi_1=0.87$,$\theta_2=2.23$, $\phi_2=0.96$, $\theta_3=1.53$, $\phi_3=0.95$, $\theta_4=0.63$, $\phi_4=1.04$ and state parameters, $k_1=k_2=\frac{1}{\sqrt{2}}$. Thus, the function $K_4$ becomes
\begin{align}
& K_4= \nonumber\\
& 2.83 + 3.84 p^6 (1 - \mu)^3 + p (-10.80 + 9.13 \mu) \nonumber\\
& + p^2 (20.74 - 36.40 \mu + 15.21 \mu^2) \nonumber\\
& + p^4 (31.92 - 94.97 \mu + 93.39 \mu^2 - 30.34 \mu^3) \nonumber\\
& + p^3 (-31 + 80.30 \mu - 64.97 \mu^2 + 15.88 \mu^3) \nonumber\\
& + p^5 (-17.27 + 53.55 \mu - 55.29 \mu^2 + 19.01 \mu^3)
\end{align}

In the Fig.\ref{DL3} we plot $K_4$ against the depolarising parameter $p$ and we see that as the memory co-efficient increases from 0 to 1, the violation of LGI can be  preserved to a considerable extent.

\section{Temporal Steering  in the presence of quantum memory channels} \label{5}

\begin{figure}[!ht]
\resizebox{9cm}{3.75cm}{\includegraphics{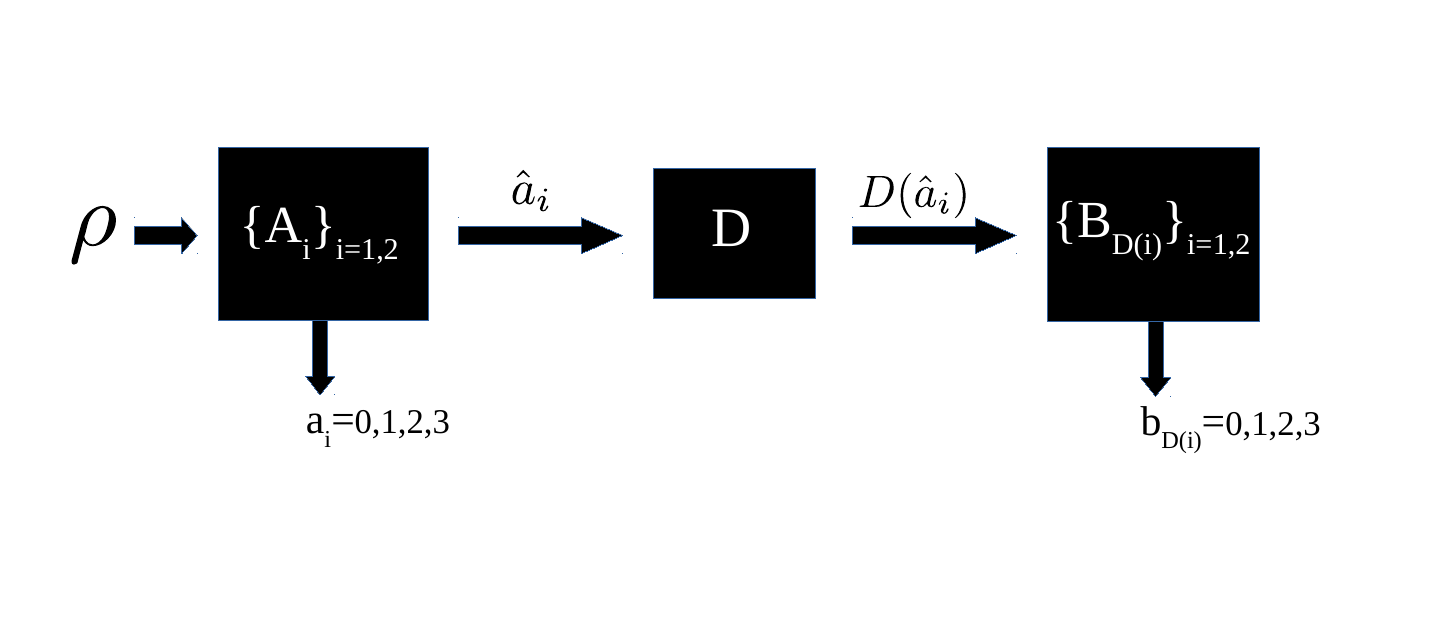}}
\caption{\footnotesize The initial state $\rho$ is first measured by Alice from a set of observables,$\lbrace A_1,A_2 \rbrace$, and then passes through the environment $D$ for a period of time ($t_B-t_A=t$), and is finally measured by Bob from a set of observables, $\lbrace B_{D(1)},B_{D(2)} \rbrace$.}
\label{Depict1}
\end{figure}

We now consider the temporal analogue of quantum steering\cite{Chen, Li}. An observer Alice measures the observable $A_i$ at time $t_A$, and sends the system to Bob through some noisy channel. For our purpose, we consider the channel to be any  of the above decoherence channels(D), {\it viz.} amplitude-damping, phase-damping channel and depolarising channels. After receiving the system, Bob measures observable $B_{D(j)}$ at a later time $t_B$ ($t_B > t_A$). Bob is restricted to measure in mutually Unbiased basis (MUB), {\it viz}., $\lbrace \sigma_x,\sigma_y,\sigma_z\rbrace$ for qubits. The outcomes of $A_i$ and $B_{D(j)}$ are $a_i$ and $b_{D(j)}$, respectively. Assuming that Alice's choice of measurements has  no influence on the state which Bob receives, one can derive a temporal steering inequality(TSI)\cite{Li} for any 4-dimensional system,  given by
\begin{eqnarray}
S_4 = \sum_{i=1}^2 \sum_{a_i=0,b_{D(i)}=a_i}^{3} P(a_i,b_{D(i)}) < \frac{3}{2}
\label{TS}
\end{eqnarray}
Violation of the above inequality infers the ability of Alice to gain information about Bob's system, or the influence of Alice's choice of observables on Bob's results. The maximum of TSI irrespective of the dimension of the measured system is $2$, which can be achieved when Alice and Bob both perform the same incompatible measurements. Unlike spatial steering, temporal steering is only one-way due to irreversibility of time. The model for testing TSI in shown  in Fig.\ref{Depict1}. 

We  consider MUB of dimension $4$ for bipartite qubit states. There are $5$ such MUBs\citep{Andreas, Andreas1, Andreas2}, which are given as follows:
\begin{align}
& \mathcal{M}_0=\lbrace \begin{pmatrix}
1\\
0\\
0\\
0\\
\end{pmatrix}, \begin{pmatrix}
0\\
1\\
0\\
0\\
\end{pmatrix}, \begin{pmatrix}
0\\
0\\
1\\
0\\
\end{pmatrix}, \begin{pmatrix}
0\\
0\\
0\\
1\\
\end{pmatrix}\rbrace, \nonumber\\
& \mathcal{M}_1= \lbrace \frac{1}{2} \begin{pmatrix}
1\\
1\\
1\\
1\\
\end{pmatrix}, \frac{1}{2} \begin{pmatrix}
1\\
1\\
-1\\
-1\\
\end{pmatrix}, \frac{1}{2} \begin{pmatrix}
1\\
-1\\
-1\\
1\\
\end{pmatrix}, \frac{1}{2} \begin{pmatrix}
1\\
-1\\
1\\
-1\\
\end{pmatrix}\rbrace, \nonumber\\
& \mathcal{M}_2= \lbrace \frac{1}{2} \begin{pmatrix}
1\\
-1\\
-i\\
-i\\
\end{pmatrix}, \frac{1}{2} \begin{pmatrix}
1\\
-1\\
i\\
i\\
\end{pmatrix}, \frac{1}{2} \begin{pmatrix}
1\\
1\\
i\\
-i\\
\end{pmatrix}, \frac{1}{2} \begin{pmatrix}
1\\
1\\
-i\\
i\\
\end{pmatrix}\rbrace, \nonumber\\
& \mathcal{M}_3= \lbrace \frac{1}{2} \begin{pmatrix}
1\\
-i\\
-i\\
-1\\
\end{pmatrix}, \frac{1}{2} \begin{pmatrix}
1\\
-i\\
i\\
1\\
\end{pmatrix}, \frac{1}{2} \begin{pmatrix}
1\\
i\\
i\\
-1\\
\end{pmatrix}, \frac{1}{2} \begin{pmatrix}
1\\
i\\
-i\\
1\\
\end{pmatrix}\rbrace, \nonumber\\
& \mathcal{M}_4= \lbrace \frac{1}{2} \begin{pmatrix}
1\\
-i\\
-1\\
-i\\
\end{pmatrix}, \frac{1}{2} \begin{pmatrix}
1\\
-i\\
1\\
i\\
\end{pmatrix}, \frac{1}{2} \begin{pmatrix}
1\\
i\\
-1\\
i\\
\end{pmatrix}, \frac{1}{2} \begin{pmatrix}
1\\
i\\
1\\
-i\\
\end{pmatrix}\rbrace
\label{MUB}
\end{align} 
The observables $A_i$ and $B_{D(j)}$ are sets of projectors corresponding to the bases given above. Among these bases, $\lbrace \mathcal{M}_0,\mathcal{M}_1,\mathcal{M}_2 \rbrace$ are separable in terms of the directions, $\lbrace ZZ,XX,YY \rbrace$, and the other two are non-separable. 

In the present analysis we apply the amplitude-damping, phase-damping and depolarising channels with strength $p$ in the time-gap, $t_A - t_B=t$. To get the maximum violation of the temporal steering inequality, we apply the same measurements for Alice and Bob each time. The joint probabilities are simply $p(a_i,b_j)=p(a_i) p(b_j|a_i)$ according to the Bayes rule. Now, we probe various channels with memory to test the TSI (Eq.\ref{TS}) in this scenario.         

\subsection{Amplitude damping channel with memory}

Let us consider that Alice measures the observable corresponding to the basis $\mathcal{M}_1$ at time $t_A$ while Bob measures the observable in the same basis $\mathcal{M}_1$ at time $t_B$. Moreover, Bob performs measurement in the basis $\mathcal{M}_2$ at time $t_B$ whenever Alice's choice of basis is $\mathcal{M}_2$ at time $t_A$. These choices make the left hand side of Eq.(\ref{TS}) maximum quantum mechanically. Now, for these measurements and the Kraus operations given in Eq.(\ref{A1},\ref{A2}) applied to the initial two-qubit state of the form of Eq.(\ref{initial}),  the quantity $S_4$ may be expressed as a function of the amplitude-damping parameter $p$ and the memory co-efficient $\mu$, as 
\begin{align}
S_4= & \frac{1}{4} \lbrace 4 + 4 \sqrt{1-p} - 2p (1 + \sqrt{1-p}) (1 - \mu) \nonumber\\
& + (1- \sqrt{1-p}) \mu\rbrace
\end{align}  
Note that $S_4$ is independent of the initial state parameters $k_1$ and $k_2$. Moreover, if one chooses any pair of MUBs from Eq.(\ref{MUB}) as observables for Alice's (and Bob's) side, the expression of $S_4$ remains the same. 

\begin{figure}[!ht]
\resizebox{6cm}{4cm}{\includegraphics{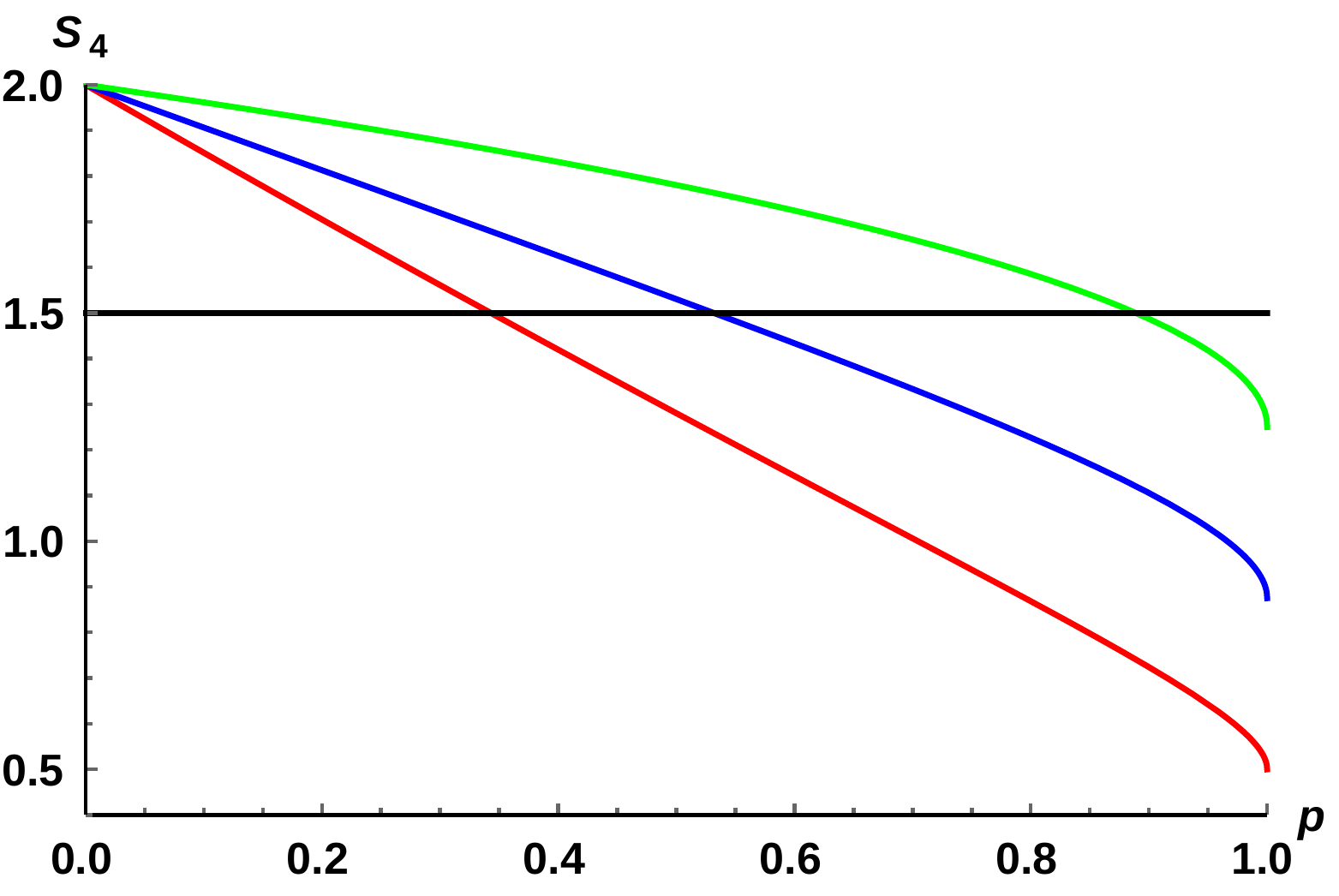}}
\caption{\footnotesize (Coloronline) $S_4$ is plotted against the amplitude-damping co-efficient $p$ for the memory strengths, $\mu=0$ (bottom red curve), $\mu=\frac{1}{2}$ (middle blue curve) and $\mu=1$ (top green curve) applying mutually unbiased measurements w.r.t. a pair of bases $\mathcal{M}_1$ and $\mathcal{M}_2$. The horizontal line represents the classical bound of temporal steering.}
\label{AT}
\end{figure} 

We plot $S_4$ against damping co-efficient $p$ in the Fig.\ref{AT}. We observe that as decoherence increases, the violation of Eq.(\ref{TS}) diminishes. However, the violation can be preserved up to significantly higher values of $p$ by increasing the memory co-efficient $\mu$.

\subsection{Phase damping channel with memory}

In a similar way, Alice and Bob both choose observables from the set of MUBs, $\lbrace \mathcal{M}_1,\mathcal{M}_2 \rbrace$. Now, using the Kraus operators expressed in Eq.(\ref{P1},\ref{P2}), and considering the bipartite pure initial state given by Eq.(\ref{initial}), we find the QM optimum of $S_4$ to be $2$. The expression for $S_4$ is given by
\begin{eqnarray}
S_4= 2 (1 - p) \lbrace 1 - p (1 - \mu)\rbrace
\end{eqnarray}
Here too, $S_4$ does not depend on the initial state, and is invariant under the choice of a pair of observables from Eq.(\ref{MUB}).

\begin{figure}[!ht]
\resizebox{6cm}{4cm}{\includegraphics{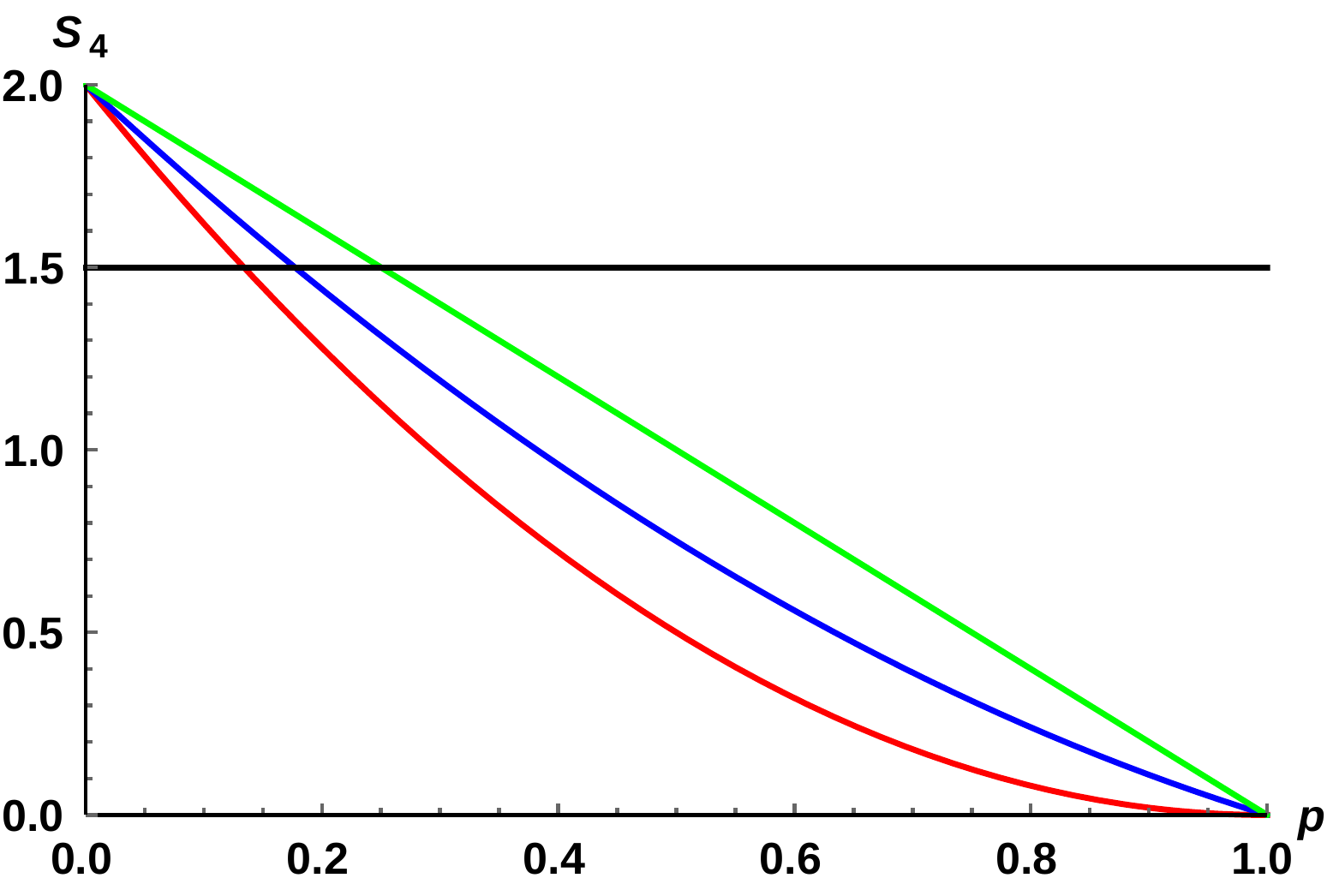}}
\caption{\footnotesize (Coloronline) A plot of $S_4$ vs the phase-damping parameter $p$ for the memory co-efficients, $\mu=0$ (bottom red curve), $\mu=\frac{1}{2}$ (middle blue curve) and $\mu=1$ (top green curve) using MUBs $\mathcal{M}_1$ and $\mathcal{M}_2$.}
\label{PT}
\end{figure} 

From Fig.\ref{PT}, we see that with increase in damping the magnitude of $S_4$ falls below the steerability bound quickly. The effect of increasing memory is able to preserve steerabilty upto slightly higher values of $p$.

\subsection{Depolarising channel with memory}

We apply the depolarising channel characterised by the Kraus operators in Eq.(\ref{D1},\ref{D2}) between the interval of Alice's and Bob's measurements, and  choose MUBs, $\lbrace \mathcal{M}_1,\mathcal{M}_2 \rbrace$  given by Eq.(\ref{MUB}) for both Alice's and Bob's measurement. These provide the QM maximum of the left hand side of TSI to be $2$. Hence, we are able to express $S_4$ as a function of the depolarising strength $p$, and the memory variable $\mu$, given by
\begin{eqnarray}
S_4= \frac{1}{9} \lbrace 18 - 15 p (2 - \mu) + 16 p^2 (1 - \mu)\rbrace
\end{eqnarray}
We note that temporal steering is unaltered for any choice of the pure initial state. Moreover, any pair of MUBs leaves the left hand side of TSI unchanged.

\begin{figure}[!ht]
\resizebox{6cm}{4cm}{\includegraphics{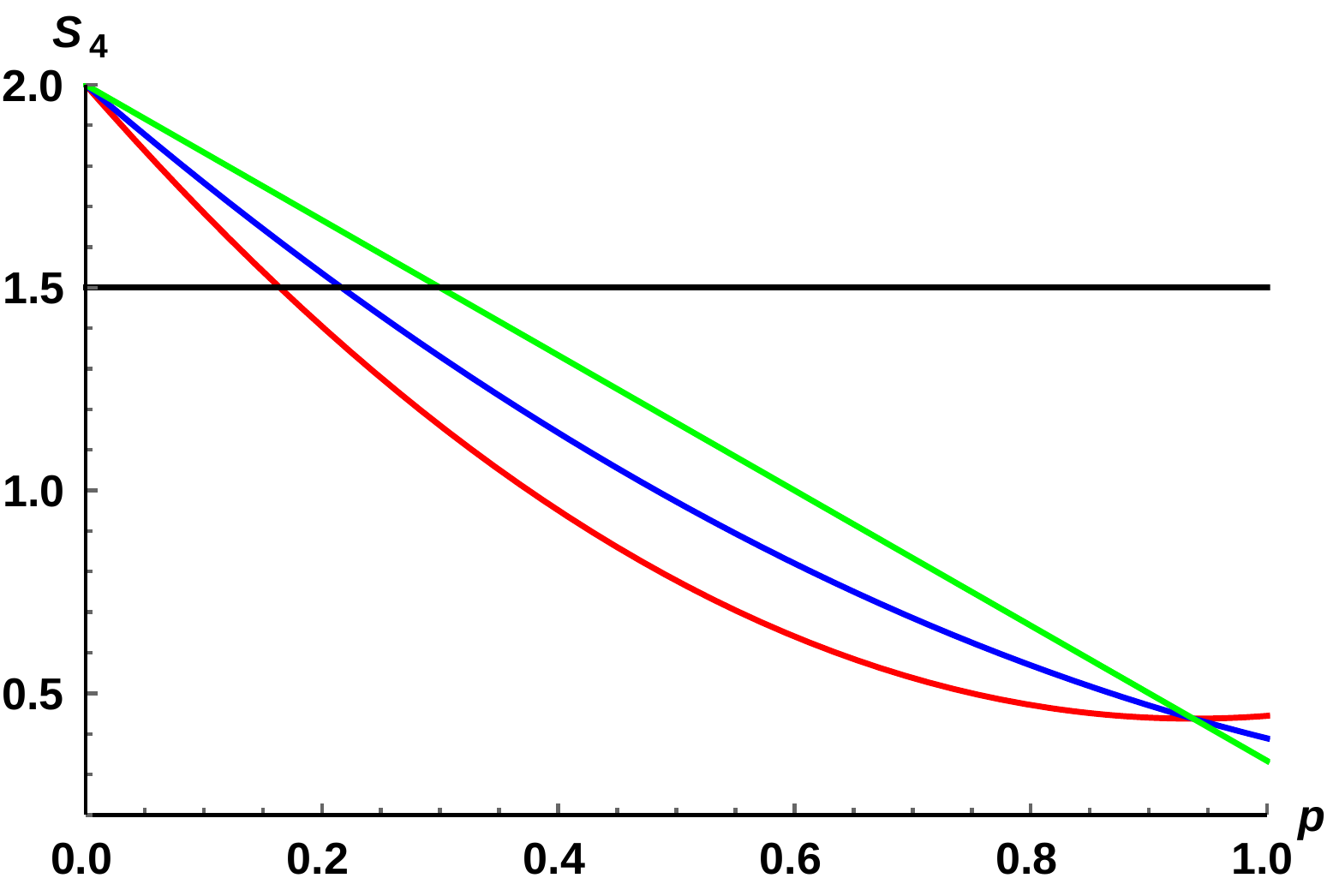}}
\caption{\footnotesize (Coloronline) Plot of $S_4$ with the depolarising strength $p$ for the values of memory co-efficients, $\mu=0$ (bottom red curve), $\mu=\frac{1}{2}$ (middle blue curve) and $\mu=1$ (top green curve) using MUBs $\lbrace \mathcal{M}_1,\mathcal{M}_2 \rbrace$.}
\label{DT}
\end{figure} 

We plot $S_4$ with the depolarising strength $p$ in Fig.\ref{DT}. It is clear from the figure that $S_4$ monotonically decreases with increasing $p$. A growing amount of memory can preserve the temporal steering correlation up to somewhat higher magnitudes of depolarization.


\section{Conclusions} \label{6}

In the present work we have investigated the dynamics of quantum temporal correlations in the presence of decoherence. Specifically, we have considered two types of temporal correlations defined respectively, by the violation of the Leggett-Garg inequality\citep{Anupam} and the temporal steering inequality\citep{Chen}, under the action of three types of decoherence given by amplitude damping, phase damping and depolarizing channels. Our motivation here has been to study to what extent the effect of decoherence diminishing the temporal correlations can be checked by employing memory or the effect of time-correlations between two successive uses of a channel\citep{Yguo}. In our analysis the initial state is taken to be a two-qubit pure entangled state subjected to three types of Bell-state measurements for probing the LGI, and measurements performed in mutually unbiased bases for studying temporal steering.

We find that in general, the effect of memory can be used to counteract the destruction of temporal correlations due to decoherence.  We show that by increasing the memory parameter, the preservation of both types of temporal correlations above their respective classical bounds ensues for higher values of the strength of decoherence.  Though this effect is observed for all the types of decoherence channels and measurements considered, there are significant quantitative differences. In particular, the magnitude of the temporal correlations diminished  due to decoherence and protected by memory varies from channel to channel depending upon the specific corresponding Kraus operator for both the LGI and TSI correlations, and also upon the specific choice of the BSM in the former case. 

Before concluding, we note that there are certain key differences in the way the dynamics of decoherence channels impacts LGI violation compared to the case of TSI violation. First, the former may depend upon the two-qubit state (Schmidt coefficients) chosen, while the latter is independent of the initial state. Secondly, if the same measurement settings are chosen by both the parties, there is no violation of LGI, whereas TSI is violated maximally in this case. Thirdly, the magnitude of LGI violation depends significantly on the particular choice of the BSM employed, while the magnitude of TSI violation is independent of the
particular type of the two-qubit MUB chosen. 

To summarize, our results show that in  spite of the above differences, both these types of temporal correlations can be preserved to various extents by the use of memory in noisy channels. Finally, it may worthwhile to pursue more such studies involving generalized noise frameworks\citep{nonmarkov}, as well as other methods\citep{Guo, Pramanik, Kim, Datta} to overcome their effects. Such investigations are essential in order to ascertain the practical viability of employing temporal correlations in quantum information processing tasks\citep{brukner'04, mal'16, shenoy'17}.

{\it Acknowledgements}: The authors thank You-Neng Guo for useful discussions. SD acknowledges financial support from DST-INSPIRE Fellowship, Govt. of India (Grant No. C/5576/IFD/2015-16).

\bibliography{Memory3}
\end{document}